\documentclass[11pt,a4paper]{article}%
\pdfoutput=1
\usepackage{jheppub}
\usepackage{amsmath,amssymb,amsfonts,wasysym}
\usepackage{mathrsfs}
\usepackage[bf,footnotesize]{caption2}
\usepackage{bbm}
\usepackage{color}
\usepackage{ulem}
\usepackage{url} 
\usepackage{latexsym,epsfig,amssymb,amsmath,slashed}

\def\({\left(}
\def\){\right)}

\def\be{\begin{equation}}
\def\ee{\end{equation}}
\def\bry{\begin{array}}
\def\ery{\end{array}}
\def\bes{\begin{subequations}}
\def\ees{\end{subequations}}
\def\bit{\begin{itemize}}
\def\eit{\end{itemize}}
\def\ben{\begin{enumerate}}
\def\een{\end{enumerate}}

 \def\be{\begin{equation}} \def\ee{\end{equation}}
\def\bea{\begin{eqnarray}} \def\eea{\end{eqnarray}}

\newcommand{\Tr}{{\rm Tr}}

\newcommand{\Dsl}{D\llap{/\kern+1.5pt}}
\newcommand{\MET}{E\llap{/\kern1.5pt}_T}

\title{\mbox{Selectron NLSP in Gauge Mediation}}
\author[a]{Lorenzo Calibbi}
\author[b]{Alberto Mariotti}
\author[c,d,e]{Christoffer Petersson}
\author[c,d]{Diego Redigolo}
\affiliation[a]{Service de Physique Th\'eorique, Universit\'e Libre de Bruxelles,\\
 C.P.~225, 1050 Brussels, Belgium}
\affiliation[b]{Institute for Particle Physics Phenomenology,
Department of Physics, Durham University,\\ DH1 3LE, United Kingdom}
\affiliation[c]{Physique Th\'eorique et Math\'ematique, Universit\'e Libre de Bruxelles,\\
 C.P.~231, 1050 Brussels, Belgium}
\affiliation[d]{International Solvay Institutes, Brussels, Belgium}
\affiliation[e]{Department of Fundamental Physics, Chalmers University of Technology,\\
412 96 G\"oteborg, Sweden}
\emailAdd{lorenzo.calibbi@ulb.ac.be} 
\emailAdd{alberto.mariotti@durham.ac.uk} 
\emailAdd{christoffer.petersson@ulb.ac.be} 
\emailAdd{diego.redigolo@ulb.ac.be}
\proceeding{\vspace{-1.65cm}
\flushright{\small IPPP/14/44\\
\small DCPT/14/88\\
ULB-TH/14-09\\}}
\abstract{We discuss gauge mediation models in which the smuon and the selectron are mass-degenerate co-NLSP, which we, for brevity, refer to as selectron NLSP. In these models, the stau, as well as the other superpartners, are parametrically heavier than the NLSP.  We start by taking a bottom-up perspective and investigate the conditions under which selectron NLSP spectra can be realized in the MSSM. We then give a complete characterization of gauge mediation models realizing such spectra at low energies. The splitting between the slepton families is induced radiatively by the usual hierarchies in the Standard Model Yukawa couplings and hence, no new sources of flavour misalignment are introduced. We construct explicit weakly coupled messenger models which give rise to selectron NLSP, while accommodating a 126 GeV MSSM Higgs mass, both within the framework of General Gauge Mediation and in extensions where direct couplings between the messengers and the Higgs fields are present. In the latter class of models, large A-terms and relatively light stops can be achieved. The collider signatures of these models typically involve multilepton final states. We discuss the relevant LHC bounds and provide examples of models where the decay of the NLSP selectron is prompt, displaced or long-lived. The prompt case can be viewed as an ultraviolet completion of a simplified model recently considered by the CMS collaboration. 
}
\begin{document} 
\maketitle
\setcounter{page}{2}
\section{Introduction}

The Large Hadron Collider (LHC) experiments, employing data from the $\sqrt{s}$=7 and 8 TeV runs, have placed considerable constraints on the strongly-interacting sector of supersymmetric (SUSY) models, with lower limits being set at about 1$\div$1.5 TeV on the masses of the gluino and the first two generations of squarks \cite{Chatrchyan:2014lfa,Aad:2014pda}, and up to around $700$ GeV for stops \cite{CMS:2013cfa,ATLAS:2013cma}.
Together with the requirements on the stop sector coming from the Higgs mass measurements \cite{Draper:2011aa}, this forces the colored states to be heavy, putting under stress the paradigm of naturalness, at least for minimal realizations of low-energy SUSY.

On the other hand, the electroweak (EW) sector of SUSY models is less constrained by direct SUSY searches and the purely EW states such as sleptons, neutralinos and charginos might in principle be significantly lighter than the rest of the spectrum. 
However, the LHC has recently started to set impressive bounds also on the EW sector, often far beyond the LEP limits.  
The present bounds for sleptons are around $300$ GeV and 500$\div$600 GeV for charginos  \cite{CMS:2013dea,Chatrchyan:2014aea,Aad:2014vma,Aad:2014nua}, and further improvements are expected from the upcoming $\sqrt{s}$=13/14 TeV run. 
We find it important to survey non-standard spectra and signatures, in order to fully exploit the LHC discovery potential in terms of EW SUSY particles.

In this paper we discuss models of gauge mediation (GM) in the 
 Minimal Supersymmetric Standard Model (MSSM) which have the non-standard property that 
the right-handed (RH) selectron and smuon are the (mass-degenerate) next-to-lightest superpartners (NLSP). The lightest SUSY particle (LSP) is the approximately massless gravitino, while the remaining superpartners, including the RH stau, are all parametrically heavier. We will refer to this 
exotic GM spectrum as the selectron NLSP scenario, in order to distinguish it from the so-called slepton co-NLSP scenario which refers to the case where the RH selectron, smuon and stau are all nearly mass-degenerate co-NLSP \cite{Ruderman:2010kj}. Throughout this paper, ``slepton" refers only  to either a selectron or a smuon, in accordance with the usual experimental separation between leptons and taus. 

We take a bottom-up perspective and explore how the selectron NLSP scenario can be realized in the framework of General Gauge Mediation (GGM) \cite{Meade:2008wd}. We will also consider extensions of GGM in which the GM messenger sector is directly coupled to the Higgs sector of the MSSM \cite{Chacko:2001km,Komargodski:2008ax,Evans:2010kd,Evans:2011bea,Evans:2012hg,Kang:2012ra,Craig:2012xp,Evans:2013kxa}. 
One virtue of the UV completions we present is that the desired hierarchies in the slepton sector are induced by the already existing flavor texture of the Yukawa couplings in the Standard Model (SM). Thus, in these setups, the selectron NLSP scenario is realized without introducing any new source of flavor misalignment in the slepton sector.\footnote{In models where an intrinsic flavor violation in the SUSY breaking mechanism is present, it is natural to get small soft masses for the light generations as a result of a mechanism of flavour violation suppression \cite{Nomura:2007ap}. Spectra with selectron NLSP were obtained in this context in \cite{Shadmi:2011hs}.}

From the collider point of view, the stable LSP gravitino generically gives rise to missing transverse energy ($\MET$). Therefore, the key role in characterizing the collider phenomenology of GM models is played by the NLSP, which decays to its SM partner and the gravitino. In this paper we will assume R-parity conservation. Depending on the SUSY breaking scale, the decay of the NLSP can be either prompt, displaced or long-lived on collider scales. 
We discuss the multilepton final states arising in the case where the NLSP decay is prompt, the charged tracks arising in the long-lived case, as well as the charged tracks ending with displaced lepton-vertices arising in the intermediate case. In the prompt case, since the dominant decay channel of the stau typically is the 3-body decay, via an off-shell Bino, to a tau, a lepton and an NLSP slepton (which subsequently decays into a lepton and a gravitino), stau pair production gives rise to the final state $2\tau+4\ell+\MET$ \cite{Ambrosanio:1997bq}.  A simplified model with stau NNLSP and selectron NLSP was recently  employed by the CMS collaboration in order to interpret the results of a multileptons search \cite{Chatrchyan:2014aea}, see also \cite{D'Hondt:2013ula} for further discussions concerning this simplified model.  The messenger models we present, which realize this spectrum at low energies, can be viewed as  possible UV completions of such a simplified model.

In GM models,
the MSSM soft masses are determined by the gauge quantum numbers of the corresponding superpartner. Even though the right-handed (RH) sleptons are mass-degenerate with the RH stau at the messenger scale, the stau mass is usually driven lighter than the first two slepton generations at low energies due to the contributions from the Yukawa interactions to renormalisation group (RG) evolution. Moreover, the lightest stau mass eigenstate can be further separated from the RH sleptons due to stau mass-mixing. 
However, a closer inspection of the MSSM RG equations (RGEs) reveals that the sign of $X_{\tau}=2\vert y_{\tau}^2\vert(m^2_{H_{d}}+m^2_{\widetilde{\tau}_{L}}+m^2_{\widetilde{\tau}_{R}})$ is crucial for determining whether the stau is driven lighter or heavier than the mass-degenerate sleptons at low energies.
In standard GM models, $X_{\tau}$ is positive along the RG flow and, as a consequence, the stau is
driven lighter than the selectron/smuon.
Instead, if $X_{\tau}$ remains negative during most of the flow, the stau is driven heavier than the selectron, thereby making possible the realization of the selectron NLSP scenario at low energies.

A negative $X_{\tau}$ implies tachyonic scalar masses at high energies and along parts of the RG flow. 
The presence of tachyons along the flow reverses the usual effect of the Yukawa interactions on the scalar masses, making heavier the scalar particles which have the largest Yukawa coupling. In this way, the appearance of a selectron NLSP at low energy is linked to the usual hierarchy among the SM Yukawa couplings $\propto m_{\tau}:m_{\mu}:m_{e}$.  

In GGM, the squared soft masses for the two Higgs doublets are equal to the left-handed (LH) slepton soft mass in the UV,  since they carry the same gauge quantum numbers: $m^2_{H_{d}}=m^2_{H_{u}}=m^2_{\widetilde{\ell}_L}$. Hence, given the definition of $X_{\tau}$, in order to realize a selectron NLSP in GGM, one needs tachyonic boundary conditions for the sleptons at the messenger scale.
 Clearly, this possibility goes beyond the minimal GM paradigm in which the squared soft masses for the sleptons are always positive at the messenger scale. Furthermore, it requires a considerable amount of gaugino mediation to push up the scalar masses at low energies such that a tachyon free spectrum is obtained.

Alternatively, if one allows the Higgs doublets to couple directly to the messengers through a superpotential interaction, i.e.~not only radiatively via gauge interactions, then
the Higgs soft masses can get new contributions in addition to the gauge mediated ones 
(we refer to this class of models as ``deflected'').
In this case, a large and negative additional soft mass contribution to the down-type Higgs doublet can  induce a selectron NLSP at low energies.\footnote{The possibility of having slepton NLSP realized within the MSSM, by allowing the soft masses of the two Higgs doublets to be independent from the soft masses of all the other sfermions, was considered in \cite{Evans:2006sj, Grajek:2013ola}.}

We show that the selectron NLSP scenario indeed is a possible low energy spectrum for GGM models with a Higgs mass at 126 GeV and with the colored sector being
significantly heavier. We also show that the selectron NLSP scenario can be realized in models with extra Higgs-messenger interactions, in which the stops (and possibly also other colored states) are kinematically accessible at the LHC. 
In particular, we study in some detail a simple model originally proposed in \cite{Evans:2011bea} as a possible way of getting the correct Higgs mass in the MSSM through the generation of a large $A_t$. Here the down-type Higgs mixes with the messengers and it acquires a tachyonic mass at tree level. This tree-level effect is enhanced for low messenger scales and therefore, this model becomes a natural UV completion of the case where the selectron NLSP is decaying promptly.  
 
The paper is organized in the following way: in Section \ref{sec:NLSP} we study the MSSM RG equations, together with the low energy constraints, and we characterize how the messenger scale soft spectrum should look like in order to realize the selectron NLSP scenario at low energies. We first outline the qualitative features of the different possibilities in GGM and in models with Higgs-messenger couplings. We then study numerically the parameter space of different models which give rise to selectron NLSP. In Section \ref{messmodel}, as a proof of principle,  we construct some explicit and minimal messenger models which realize this scenario. Finally in Section \ref{sec:collider} we discuss the collider signatures of spectra with selectron NLSP treating separately the cases in which the NLSP decay is prompt, displaced or long-lived on collider scales. 
 
\section{Selectron NLSP scenario from the RG-flow}\label{sec:NLSP}

In this section we discuss the RGEs for the MSSM soft masses and the low-energy constraints coming from the requirements of EW symmetry breaking (EWSB) and the absence of tachyons. The purpose is to characterize the parameter space that gives rise to the selectron NLSP scenario at the EW scale. In particular, our aim is to determine the minimal requirements on the soft masses, at the messenger scale, which are necessary in order to get the stau heavier than the selectron/smuon. 
Note that we distinguish between sleptons \mbox{$\widetilde{\ell}=\widetilde{e}, \widetilde{\mu}$} and stau $\widetilde{\tau}$, as well as between leptons $\ell=e,\mu$ and  tau $\tau$. 

A necessary ingredient - though not sufficient - to realise the selectron NLSP scenario is the following condition on the low energy soft SUSY-breaking mass terms:

\begin{equation}
 \Delta^2_{R,L} \equiv\left(m_{\widetilde{\tau}_{R,L}}^2-m_{\widetilde{\ell}_{R,L}}^2 \right)>0,
 \end{equation} 
which can be mapped into specific conditions on the soft masses at the messenger scale by solving the MSSM RGEs. The splitting between the stau and the selectron gauge eigenstates can easily be derived from $\Delta^2_{R,L}$ in the case where it is small compared to the slepton masses: $m_{\widetilde{\tau}_{R,L}}-m_{\widetilde{\ell}_{R,L}} \approx  \tfrac{\Delta^2_{R,L}}{2 m_{\widetilde{\ell}_{R,L}}}$. 

The RGEs for the soft masses of the LH and RH sleptons and stau are, at one loop, given by
\bea
16 \pi^2 \frac{d}{dt} m_{\widetilde{\ell}_L/\widetilde{\tau}_L}^2 &=& X_{\ell/\tau}- 6 g_2^2 |M_2|^2-\frac{6}{5} g_1^2 |M_1|^2-\frac{3}{5} g_1^2 S\, , \label{eq:mL}\\
16 \pi^2 \frac{d}{dt} m_{\widetilde{\ell}_R/\widetilde{\tau}_R}^2 &=& 2 X_{\ell/\tau}-\frac{24}{5} g_1^2 |M_1|^2+\frac{6}{5} g_1^2 S \, ,
\label{eq:mE}
\eea
where the contributions induced by the Yukawa couplings enter via the combination
\be
X_{\ell/\tau}\equiv 2 |y_{\ell/\tau}^2| (m_{H_d}^2+m_{\widetilde{\ell}_L/\widetilde{\tau}_L}^2+m_{\widetilde{\ell}_R/\widetilde{\tau}_R}^2)\, ,
\ee
and where
\be
S\equiv m_{H_u}^2 - m_{H_d}^2+\Tr \left[m_{\widetilde{Q}}^2-2 m_{\widetilde{u}}^2+m_{\widetilde{d}}^2-m_{\widetilde{\ell}_L}^2+m_{\widetilde{\ell}_R}^2 \right]\,,\label{FI}
\ee
where the trace Tr is taken over the flavour indices of the MSSM soft mass matrices, $m_{\widetilde{Q}}^2$, $m_{\widetilde{u}}^2$, $\cdots$. 

If we neglect the contributions from the (small) lepton Yukawa couplings, we see that the RG evolution for the difference between the stau and slepton soft masses is determined by $X_\tau$:
\be
\label{RGDelta}
16 \pi^2 \frac{d}{dt}  \Delta^2_{R,L} = c_{R,L} X_{\tau} = 2 c_{R,L} |y_{\tau}^2| (m_{H_d}^2+m_{\widetilde{\tau}_L}^2+m_{\widetilde{\tau}_R}^2)\, ,
\ee
where $c_{R}=2c_{L}=2$. In leading-log approximation, the solution to Eq.~\eqref{RGDelta} reads:
\be
\label{eq:LL}
 \Delta^2_{R,L} \approx 
 -c_{R,L} \frac{m_{\tau}^2\tan^2\beta}{8\pi^2 v} (m_{H_d}^2+m_{\widetilde{\ell}_L}^2+m_{\widetilde{\ell}_R}^2)
 \ln\left(\frac{M}{M_{S}}\right)\,,
\ee
where $M$ and ${M_{S}}$ are the messenger scale and a typical low-energy soft mass, respectively, and where we have assumed flavour blind soft terms at the scale $M$ by imposing $m_{\widetilde{\ell}_L}=m_{\widetilde{\tau}_L}$ and  $m_{\widetilde{\ell}_R}=m_{\widetilde{\tau}_R}$. 
Here, the soft masses on the RHS of \eqref{eq:LL} take the values they have at the messenger scale and we neglect their running. This approximation allows us to make some useful rough estimates, but it is clearly not accurate in the case where the running of the slepton masses and/or of the down-type Higgs mass is non-negligible.

\begin{figure}[t]
\begin{center}
\includegraphics[width=0.45\textwidth]{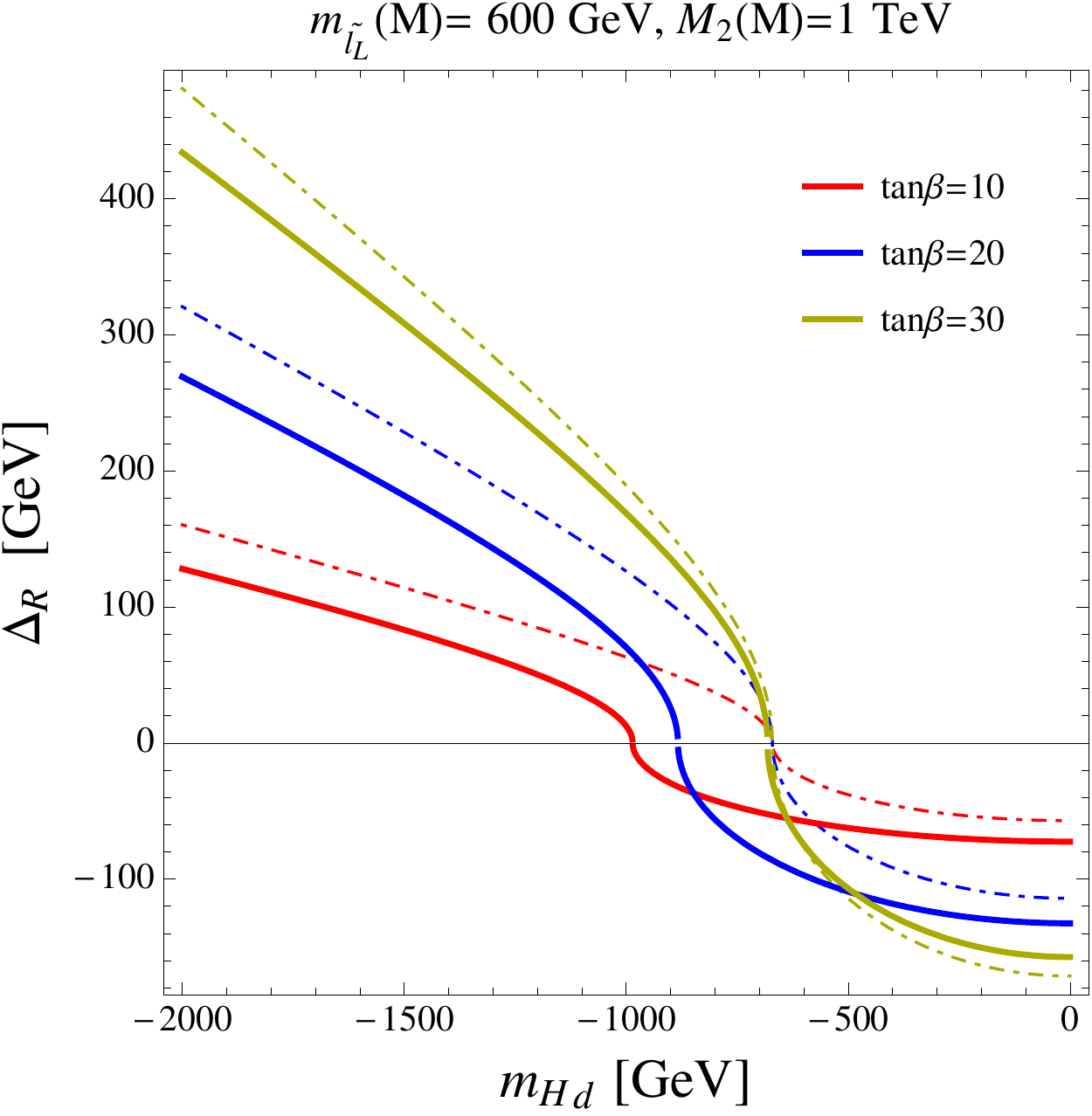}
\hspace{1cm}
\includegraphics[width=0.45\textwidth]{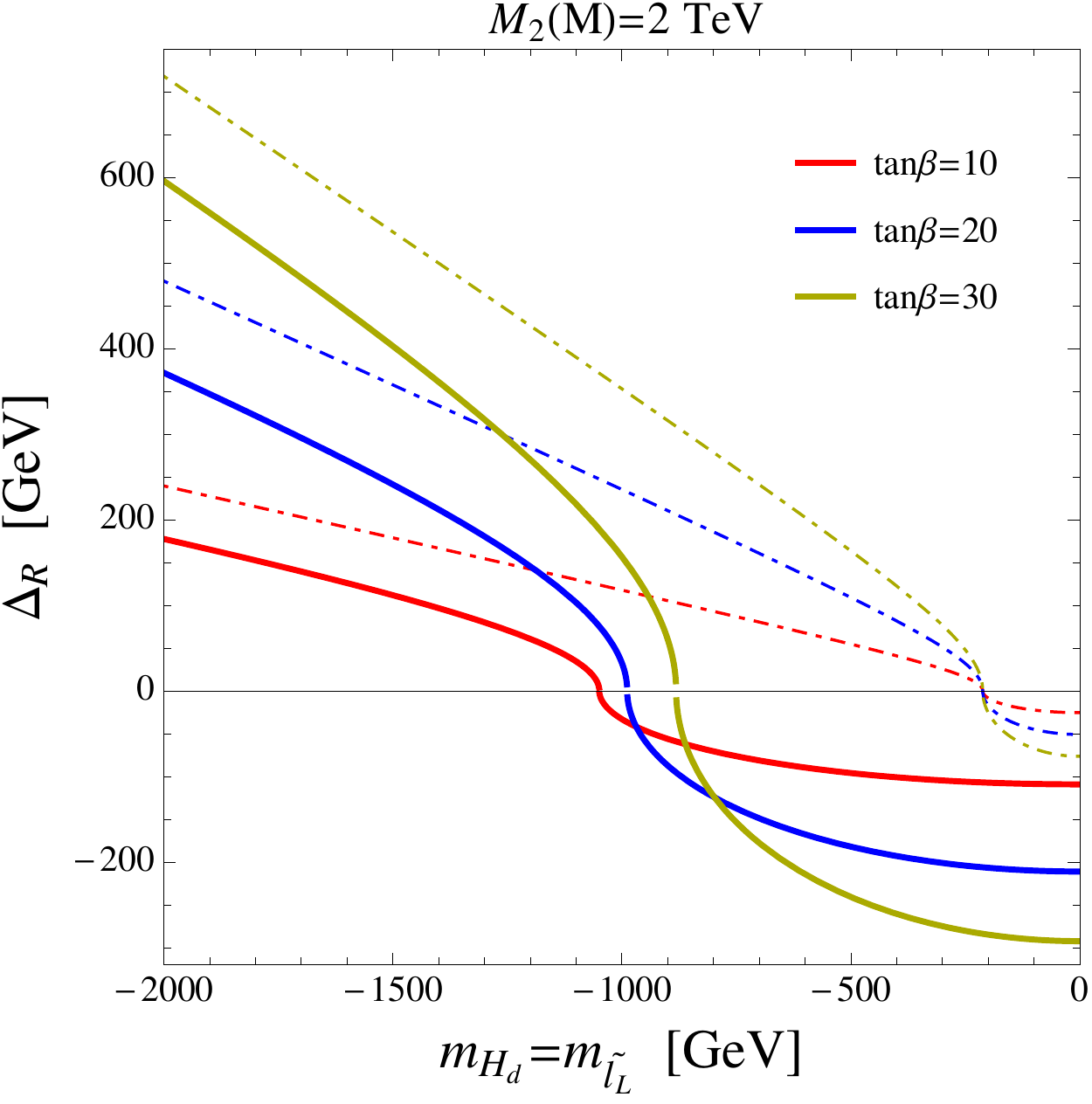}
\caption{RH stau-selectron mass splitting for different values of $\tan\beta$. The dashed lines are leading-log estimates based on Eq.~(\ref{eq:LL}). We define $\Delta_R\equiv \Delta_R^2/\sqrt{|\Delta_R^2|}$ and 
$m_{\tilde f}\equiv m_{\tilde f}^2/\sqrt{|m_{\tilde f}^2|}$.
The solid lines are exact solutions of the 1-loop RG-equations. The messenger scale is fixed to $M=10^{15}$ GeV and $M_S=1$ TeV. The two cases differ only for the value of $M_2$ and $m_{\widetilde{\ell}_L}$ which are reported in the figures. The rest of the soft spectrum at the scale $M$ is fixed to $m_{\widetilde{\ell}_R}=300\text{ GeV}$, $m_{\text{squarks}}=2 \text{ TeV}$ and $M_{3}=2 \text{ TeV}$. \label{fig:LL-est}}
\end{center}
\end{figure}

In minimal GM models, where all the three (squared) soft masses on the RHS of Eqs.~(\ref{RGDelta}, \ref{eq:LL}) are positive 
at the messenger scale, $X_{\tau}>0$, and the two stau soft masses are driven smaller than the slepton masses at low energies. 
Instead, from 
Eqs.~(\ref{RGDelta}, \ref{eq:LL}) we can also conceive the possibility of realizing situations where $X_{\tau}<0$ such that the stau masses are driven heavier than the slepton masses at low energies. 
This is the effect we need in order to realize the selectron NLSP scenario.
As is manifest in Eqs.~(\ref{RGDelta}, \ref{eq:LL}),
the key ingredient is the presence of tachyonic masses for ${H_d}$ and/or ${\tilde \ell}_{R,L}$ along the RG flow,
sufficiently large to render $X_{\tau}$ negative.

In Figure \ref{fig:LL-est} we show both the leading-log estimates and the exact 1-loop solution for the mass splitting $\Delta_R\equiv \Delta_R^2/\sqrt{|\Delta_R^2|}$ as a function of the high-energy values of $m_{H_d}$ (left) and 
$m_{{\ell}_L}=m_{H_d}$ (right). The two plots correspond to the two prototypical spectra that we will study. Figure \ref{fig:LL-est} (left) corresponds to deflected spectra where the splitting $\Delta_{R,L}$ is triggered by $m_{H_d}$ alone and where there is no need for large values of $M_2$. Instead, Figure \ref{fig:LL-est} (right), where $m_{H_d}=m_{\widetilde{\ell}_L}$ and where $M_2$ is always sizeable, corresponds to GGM spectra.  

Comparing the two plots in Figure \ref{fig:LL-est}, we see that the leading-log estimate predicts the splitting effect in the right plot to be a factor of $\sqrt{2}$ larger than in the left plot.  However, it turns out that this approximation is accurate only for small values of $M_2$. In fact, by increasing $M_2$, the gaugino mediation effect from the Wino on the LH sleptons and on the down-type Higgs becomes relevant and, as a consequence, greater tachyonic values for $m^2_{H_d}/m^2_{\widetilde{\ell}_L}$ are needed in order to get the same splitting effect. Note that the value of the squark masses controls the running of $m^2_{H_d}$ and determines the splitting of the turning points for the different curves, which is not captured by the leading-log approximation.  

We are interested in spectra in which the RH sleptons are co-NLSP.\footnote{Another interesting option could be to have the LH sleptons and their corresponding sneutrinos as co-NLSP. Many of the considerations of this paper can be easily adapted to this case, which we leave for future studies.} 
In order for the RH sleptons to be co-NLSP they should be lighter than all the other sparticles and, in particular, lighter than the physical mass of the lightest stau. In order to take into account the left-right mixing we should consider the mass matrices for the sleptons/staus, which are given by, 
  \be
\left(
\begin{array}{cc}
m_{\widetilde{\ell}_L/\widetilde{\tau}_L}^2 & -m_{\ell/\tau}\,\mu \tan \beta \\
-m_{\ell/\tau}\,\mu \tan \beta & m_{\widetilde{\ell}_R/\widetilde{\tau}_R}^2
\end{array}
\right)\label{staummatrix}\,,
\ee
where we have neglected the contributions from the corresponding $A$-terms, as well as the contributions that arise upon the EW symmetry breaking. For the sleptons, since the off-diagonal entries are negligible, we will denote the two mass eigenvalues also by $m^2_{\widetilde{\ell}_R}$ and $m^2_{\widetilde{\ell}_L}$. For the staus, since the off-diagonal entries can be relevant  - especially for large values of $\mu \tan\beta$ - we are interested in the smallest stau mass eigenvalue, given by 
 \be
m_{\widetilde \tau_1}^2= \frac{1}{2} \left(
m_{\widetilde{\tau}_L}^2+m_{\widetilde{\tau}_R}^2-\sqrt{(m_{\widetilde{\tau}_L}^2-m_{\widetilde{\tau}_R}^2)^2+ 4 m_{\tau}^2 \mu^2 \tan^2 \beta} \right)\,,
\label{tau1}
\ee 
where all the soft masses are evaluated at the EW scale. The selectron NLSP scenario requires $m^2_{\widetilde{\ell}_R}< m_{\widetilde{\tau}_1}^2$ at the EW scale.  
\begin{figure}[t]
\begin{center}
\includegraphics[width=0.4\textwidth]{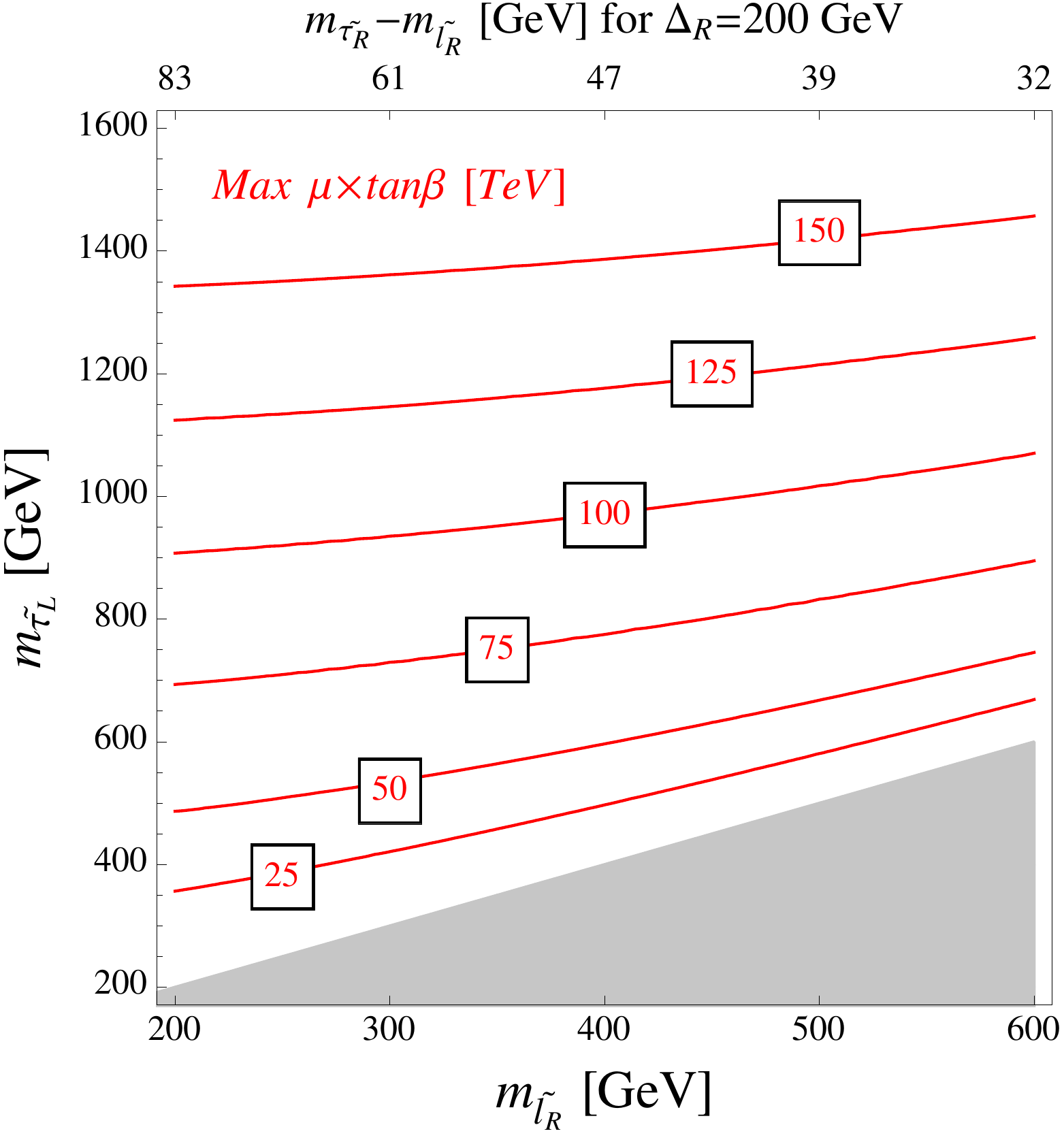}
\hspace{1cm}
\includegraphics[width=0.4\textwidth]{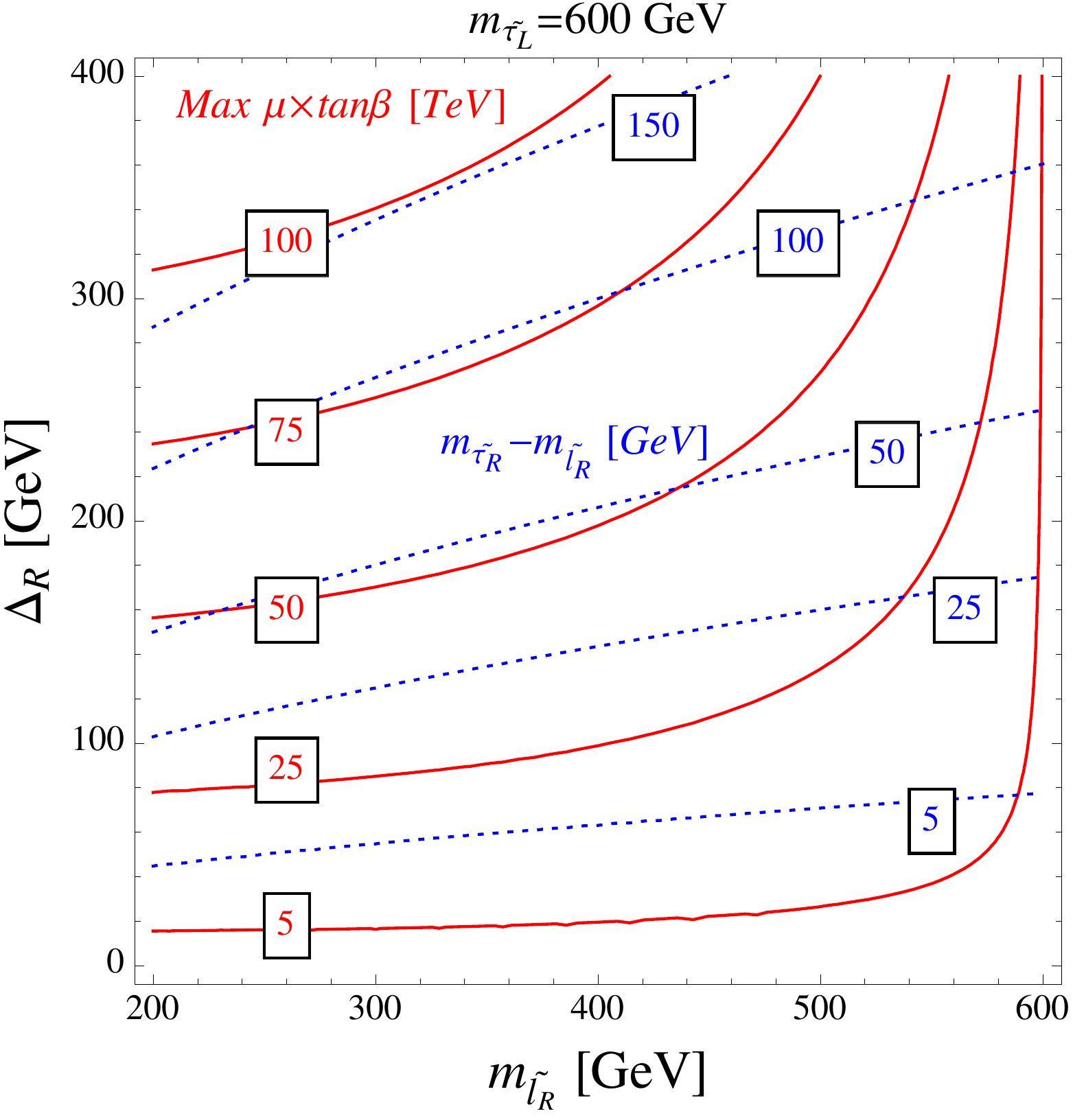}
\includegraphics[width=0.4\textwidth]{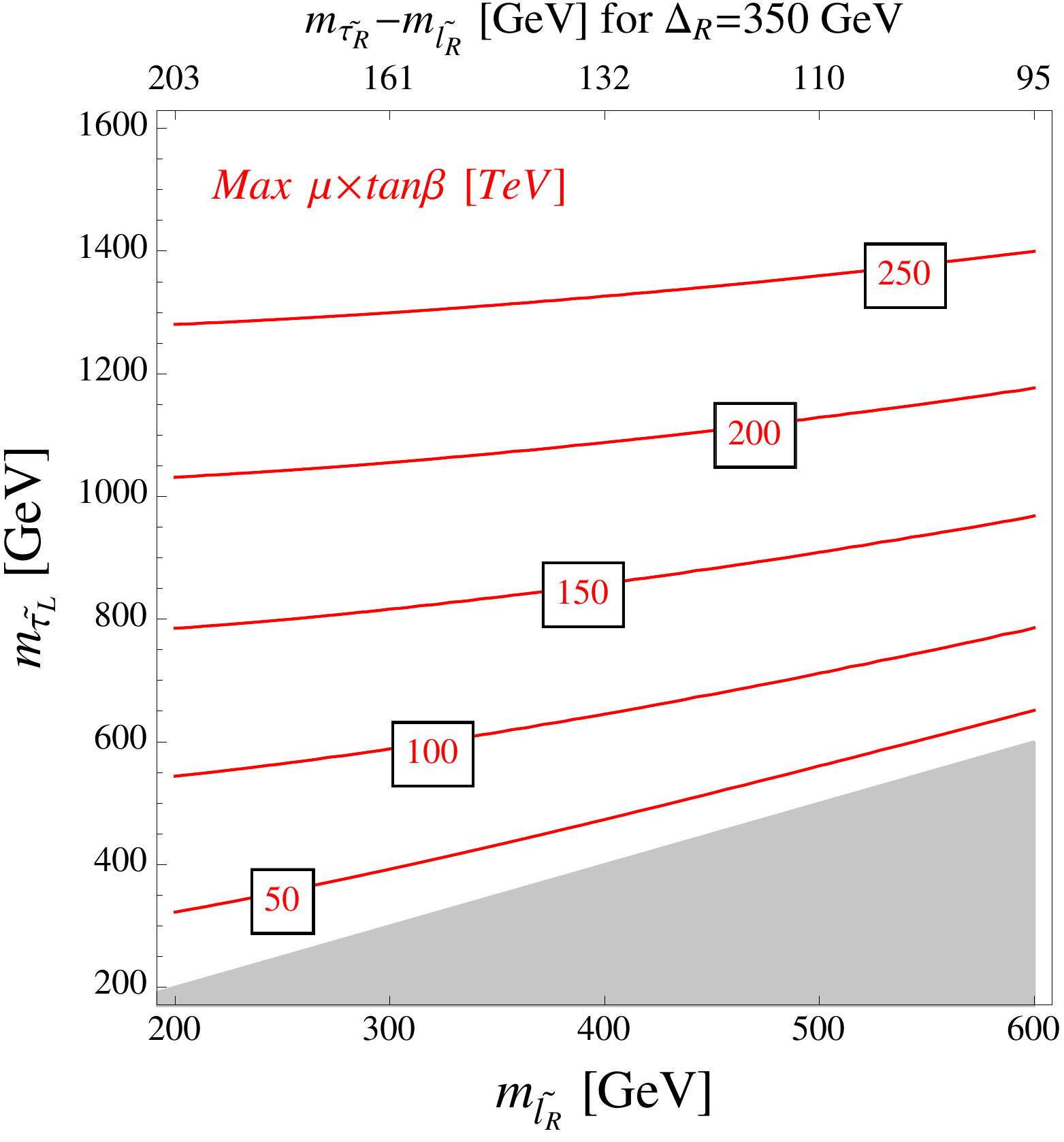}
\hspace{1cm}
\includegraphics[width=0.4\textwidth]{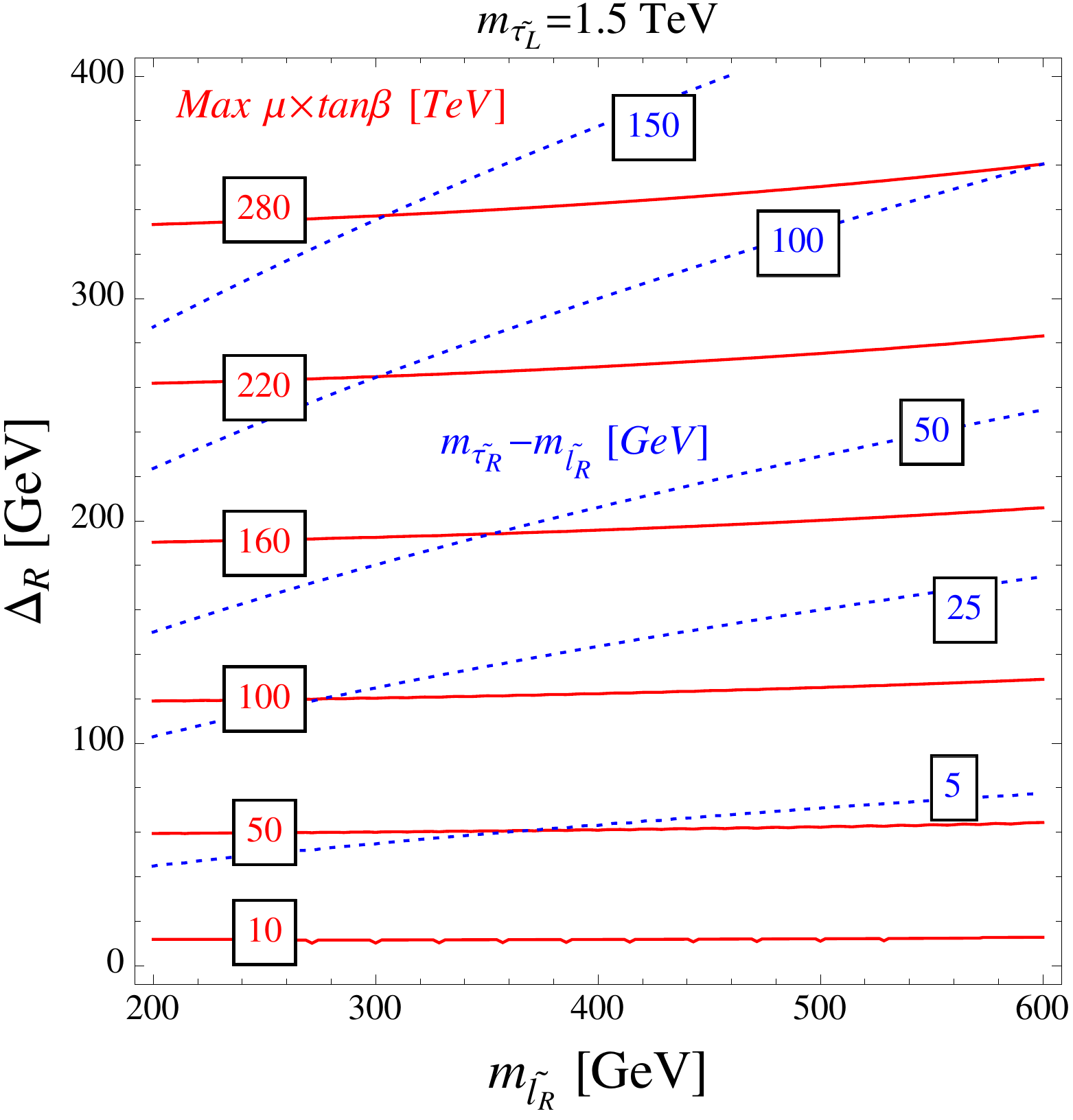}

\caption{\label{fig:mutb} Contours of the upper bounds on $\mu\times \tan\beta$ (in TeV) for different choices of the parameters of the matrix \eqref{staummatrix}. As before, $\Delta_R\equiv \Delta_R^2/\sqrt{|\Delta_R^2|}$.} 
\end{center} 
\end{figure} 

Since the tau Yukawa coupling in \eqref{RGDelta} is enhanced at large values of $\tan\beta$, it is possible to increase the separation between the RH stau and slepton by increasing $\tan\beta$. However, from \eqref{staummatrix} we see that $\tan\beta$ can not be too large since the mixing in the stau mass matrix then increases as well -- precisely how much also depends on the values of the mass parameters in \eqref{staummatrix} -- and consequently, the lightest stau, whose mass is given in \eqref{tau1}, is pushed lighter.
In other words, for a given configuration of the parameters, we will have an upper bound on $\mu\tan\beta$ from the requirement that the LR stau mixing can not be too large in order to realize $m_{\widetilde{\ell}_R}< m_{\widetilde{\tau}_1}$.
The dependence of such a bound on the slepton masses and the splitting parameter $\Delta_R$ is shown in Figure \ref{fig:mutb}. 
As we can see, this bound can be quite stringent, especially for a light LH stau. Interestingly, it can be translated into an upper bound on $\mu$, i.e.~on the Higgsino masses, taking into account that at least moderate values of  $\tan\beta$ are typically needed in order to radiatively generate a sizeable mass splitting $\Delta_R$, 
as is shown in Figure \ref{fig:LL-est}.
As we can see from Figure \ref{fig:mutb}, the bound gets significantly relaxed if we allow for a heavy LH stau.
In this figure, we also show the exact dependence of $m_{\widetilde{\tau}_{R}}-m_{\widetilde{\ell}_{R}}$ on the values of $\Delta_R$ and $m_{\widetilde{\ell}_R}$, and we see how the splitting decreases as we increase the mass of the RH sleptons, at fixed $\Delta_R$.

Let us now discuss how these two features of the selectron NLSP scenario, i.e.~the necessity of having tachyonic masses for ${H_d}$ and/or ${\tilde\tau}_{R,L}$ and to minimize the left-right mixing in the stau mass matrix, can be realized in different classes of models. In particular we are going to consider GGM models in Section \ref{sec:GGM} and models with deflections for the soft masses in the Higgs sector in Section \ref{sec:Deflection}. 

Each section will be organized as follows: we first give a brief summary of the structure of the parameter space. We then give some qualitative understanding of the RG-flow effects for points with selectron NLSP. In order to do that we solve the 1-loop RG-equations semi-analytically imposing the EWSB conditions at tree level.\footnote{We thank Simon Knapen and David Shih for sharing with us a private Mathematica code in which this is computed.} We then complement our analysis with a full numerical scan of the parameter space with {\tt SOFTSUSY 3.3.9} \cite{Allanach:2001kg}, taking into account low-energy threshold corrections and 2-loop effects. The full numerical approach is going to confirm our qualitative understanding and to realize selectron NLSP scenarios with $m_h=126\pm3 \text{ GeV}$.

\subsection{General Gauge Mediation}\label{sec:GGM}
The General Gauge Mediation (GGM) framework  consists of a hidden sector that breaks SUSY spontaneously and a visible sector that we choose to be the MSSM. The decoupling limit between the two sectors is achieved when all the SM gauge interactions are switched off: $g_{i}\to0$, for $i=1,2,3$. Consequently, the parameter space is defined at the messenger scale $M$ by two independent sum-rules $\text{Tr}(Ym^2_{\widetilde{f}})=0$ and $\text{Tr}((B-L)m^2_{\widetilde{f}})=0$, which follows from the two non-anomalous symmetries of the MSSM. Using these two relations we can write two of the five MSSM soft terms in terms of the others:  
\begin{align}
m^2_{\widetilde{u}}(M)&= m^2_{\widetilde{Q}}(M)-m^2_{\widetilde{\ell}_L}(M)+\frac{2}{3}m^2_{\widetilde{\ell}_R}(M)\ , \\
m^2_{\widetilde{d}}(M)&= m^2_{\widetilde{Q}}(M)-m^2_{\widetilde{\ell}_L}(M)+\frac{1}{3}m^2_{\widetilde{\ell}_R}(M)\ .
\end{align}
The independent GGM soft parameters at the messenger scale are then reduced to three complex gaugino masses $(M_1,\ M_2,\ M_3)$, which are here taken to be real, and three real sfermion masses $(m_{\widetilde{Q}},\ m_{\widetilde{\ell}_L},\ m_{\widetilde{\ell}_R})$. These soft masses can be written as 
\begin{align}
& M_{i}(M)=\frac{g_{i}^2(M)}{(4\pi)^2}\Lambda_{G_{i}}\, , \label{gmass}\\
& m^2_{\widetilde{f}}(M)=2\sum_{i=1}^{3}C_{\widetilde{f}_{i}}\frac{g^4_{i}(M)}{(4\pi)^4}\Lambda_{S_{i}}^2\, .\label{sfmass}
\end{align}
where $C_{\widetilde{f}_{i}}=(3/5Y^2,3/4,4/3)$ is the quadratic Casimir for the representation $\widetilde{f}$ under the $i^{th}$ gauge group of the SM, with the GUT normalization for $g_1$. $\Lambda_{G_{i}}$ and $\Lambda_{S_{i}}$ are model-dependent functions of the SUSY-breaking scales of the hidden sector and of the characteristic UV scale $M$, which we take to be unique. 

In the Higgs sector, the soft masses for the two doublets are fixed to be equal to the soft mass for the left-handed sleptons:
\begin{align}
m^2_{H_{d}}=m^2_{H_{u}}=m^2_{\widetilde{\ell}_L}. 
\label{eq:higgses}
\end{align}
If we take $\mu$ to be a free parameter in the superpotential, independent of the SUSY breaking mechanism,
then GGM sets
\begin{align}
B_{\mu}=0
\label{eq:Bmu}
\end{align}
at the messenger scale. 
Moreover, the A-terms are always suppressed in gauge mediation and can be set to zero at the messenger scale. 
The GGM parameter space is then determined by $6+2+1$ parameters, where 6 parameters describe the soft masses for gauginos and sfermions, 2 parameters characterise the EWSB $(\mu$ and $\tan\beta)$ and 1 is simply the messenger scale $M$, which sets the length of the RG-flow.  

\begin{figure}[t]
\begin{center}
\includegraphics[width=0.48\textwidth]{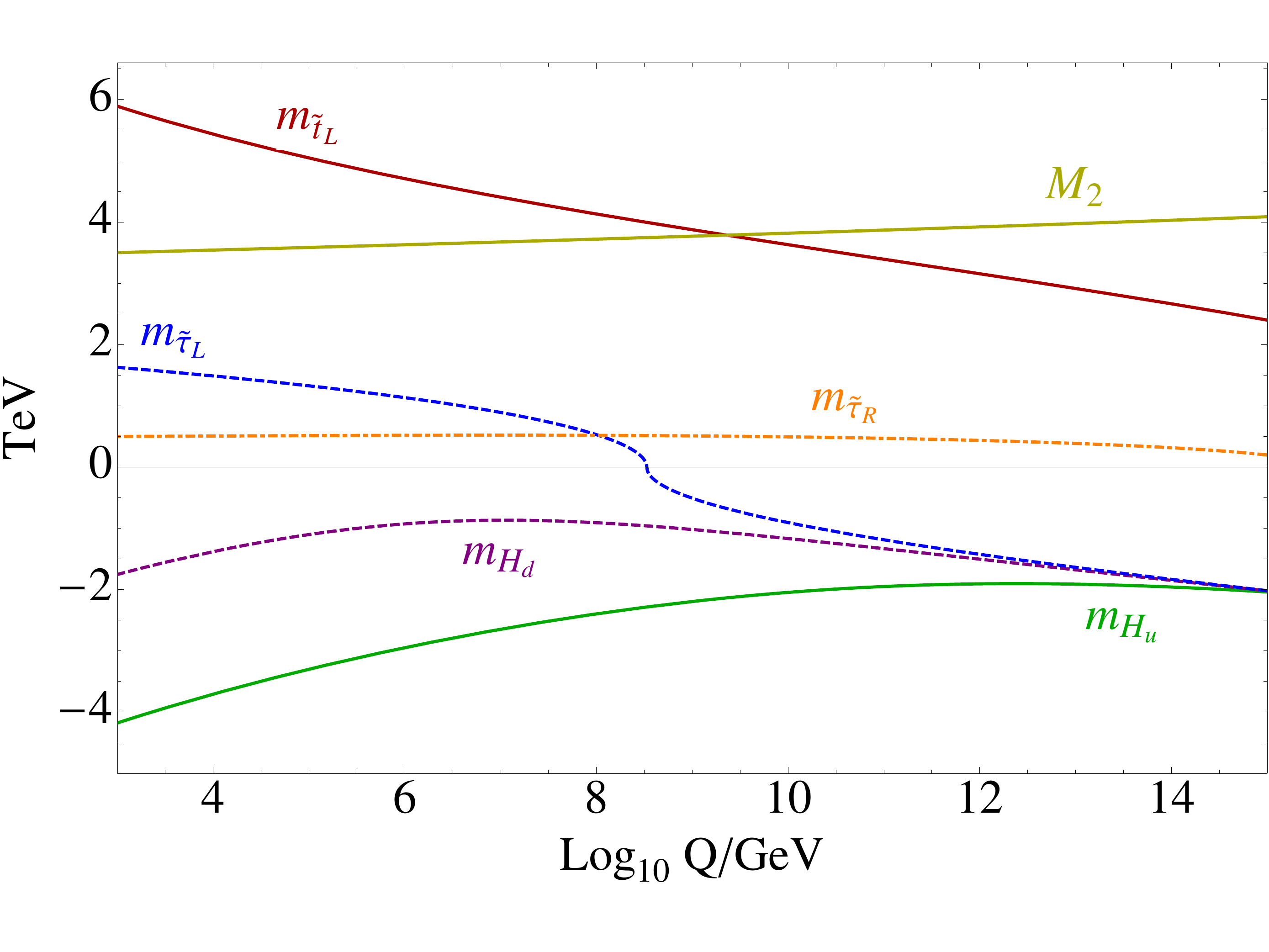}
\hspace{10pt}
\includegraphics[width=0.48\textwidth]{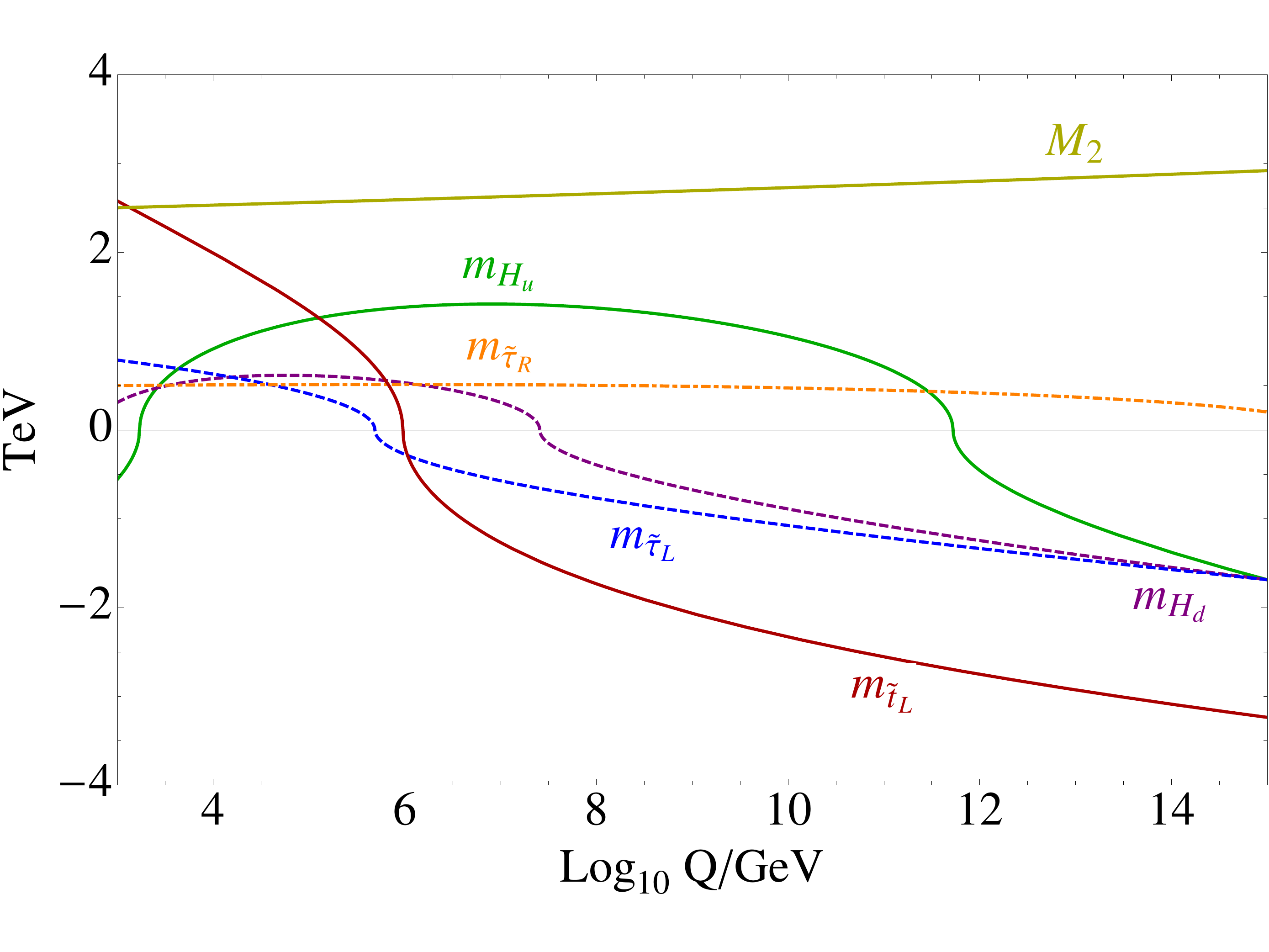}
\caption{Examples of RG flows of the soft masses from the messenger scale $M$ to $M_{S}=\sqrt{m_{\tilde{t}_L} m_{\tilde{t}_R}}$ for points with selectron NLSP in GGM scenarios. The scalar masses are defined $m_{\tilde f}\equiv m^2_{\tilde f}/\sqrt{|m^2_{\tilde f}|}$. Two GGM configurations are shown with $M=10^{15}\text{ GeV}$ and $\tan\beta=30$. In both the examples $\Delta_R=200\text{ GeV}$, $m_{\widetilde{\tau}_R}=500\text{ GeV}$ and $M_{1}=550 \text{ GeV}$.  \label{fig:RGflowsExample}  } 
\end{center}
\end{figure}

Because of the condition \eqref{eq:higgses}, the easiest way to realize the selectron NLSP scenario is to have a tachyonic mass for the left-handed sleptons at high energy, resulting in a negative $X_{\tau}$.  A sufficient amount of gaugino mediation, in particular a heavy Wino, can then drive the left-handed slepton mass positive at low energies, and even heavier than the right-handed ones. This effect is displayed in both of the RG flow examples shown in Figure \ref{fig:RGflowsExample}, where we have chosen $\Delta_R=200\text{ GeV}$ and $m_{\widetilde{\tau}_R}=500\text{ GeV}$
at low energy, which correspond to a splitting  of $m_{\widetilde{\tau}_{R}}-m_{\widetilde{\ell}_{R}}\approx40\text{ GeV}$, as can be derived from Figure \ref{fig:mutb}.\footnote{In order to get a light RH selectron NLSP at low energy we may need the RH slepton mass to start tachyonic at the messenger scale in order to counteract the effect of gaugino mediation controlled by the Bino mass. Such a tachyonic start also helps getting the desired splitting effect between the stau and the sleptons, as can be seen from Eq.~\eqref{RGDelta}.} 

The two examples in Figure \ref{fig:RGflowsExample} are distinguished by the behaviour of the squarks along the flow. In the left panel, the squark masses are positive at the messenger scale and, as a consequence, $m^2_{H_u}$ and $m^2_{H_d}$ are driven negative along the flow. Since $m^2_{H_u}$ is already forced to start tachyonic because of the GGM condition \eqref{eq:higgses}, which set it equal to $m^2_{\widetilde{\ell}_{L}}$, this scenario is characterized by a large $\mu$, which is fixed to be $\vert\mu\vert^2\approx-m_{H_u}^2$ by the EWSB condition. In the left panel of Figure \ref{fig:RGflowsExample} we get $\mu\approx4\text{ TeV}$ but the selectron NLSP scenario is still possible since $M_2$ is large enough to make the LH sleptons heavy ($m_{\widetilde{\tau}_L}=1.6\text{ TeV}$) and thereby counteract the mixing effects in the stau mass matrix, as can be seen in \mbox{Figure \ref{fig:mutb}.} 

A possibility of getting a small $\mu$ and selectron NLSP in GGM is depicted in the right panel of Figure \ref{fig:RGflowsExample}, where $\mu\approx600 \text{ GeV}$. The idea is to start with tachyonic masses for the squarks at the messenger scale, which can then be driven positive along the flow by gluino mediation. In this way, the usual effect of the stops on $m^2_{H_u}$ is partially reversed since $m^2_{H_u}$ is pulled up until the scale for which the stop masses become non-tachyonic again. Analogous spectra have been proposed as possible ways to minimize the tuning in GGM and getting large A-terms to enhance the MSSM Higgs mass \cite{Dermisek:2006ey,Draper:2011aa}. In order to get selectron NLSP, this spectrum is quite natural since it is the only way of starting with a tachyonic $m_{\widetilde{\ell}_L}$ without getting a large $\mu$ at the EW scale, hence automatically minimizing the mixing effects in the stau mass matrix (we get $m_{\widetilde{\tau}_L}\approx 800\text{ GeV}$ in our benchmark). Therefore, we expect the selectron NLSP scenario to be a possible spectrum in the GGM scenarios proposed in \cite{Dermisek:2006ey,Draper:2011aa}. 

In principle one can envisage a third possibility of getting the selectron NLSP scenario in GGM by starting with a fully non-tachyonic spectrum and triggering a negative squared mass for $H_{d}$ (large enough to make $X_{\tau}$
negative) via the RG-flow. The effect can be understood from the following RG equation:
\begin{equation}
16\pi^2\frac{d}{dt}\left(m^2_{H_{d}}-m^2_{\widetilde{\tau}_{L}}\right)=3X_{b}=6\vert y_b\vert^2\left(m^2_{H_{d}}+m^2_{\widetilde{Q}_3}+m^2_{\widetilde{d}_3}\right)\, .
\end{equation}
From this equation, we see that when the squark masses are heavy, $X_{b}>0$, and the RGEs drive $m^2_{H_{d}}<m^2_{\widetilde{\tau}_L}$. This effect becomes relevant  for large values of $\tan\beta$ because it is controlled by $y_{b}$. However, heavy squark masses would also induce a very large $\mu$, which would enhance the stau left-right mixing term. Getting a selectron NLSP is then a matter of balancing these two effects. We checked numerically that this indeed is feasible, but it requires very fine-tuned spectra. For this reason we do not consider this possibility in what follows.  

\begin{figure}[t]
\begin{center}
\includegraphics[width=0.5\textwidth]{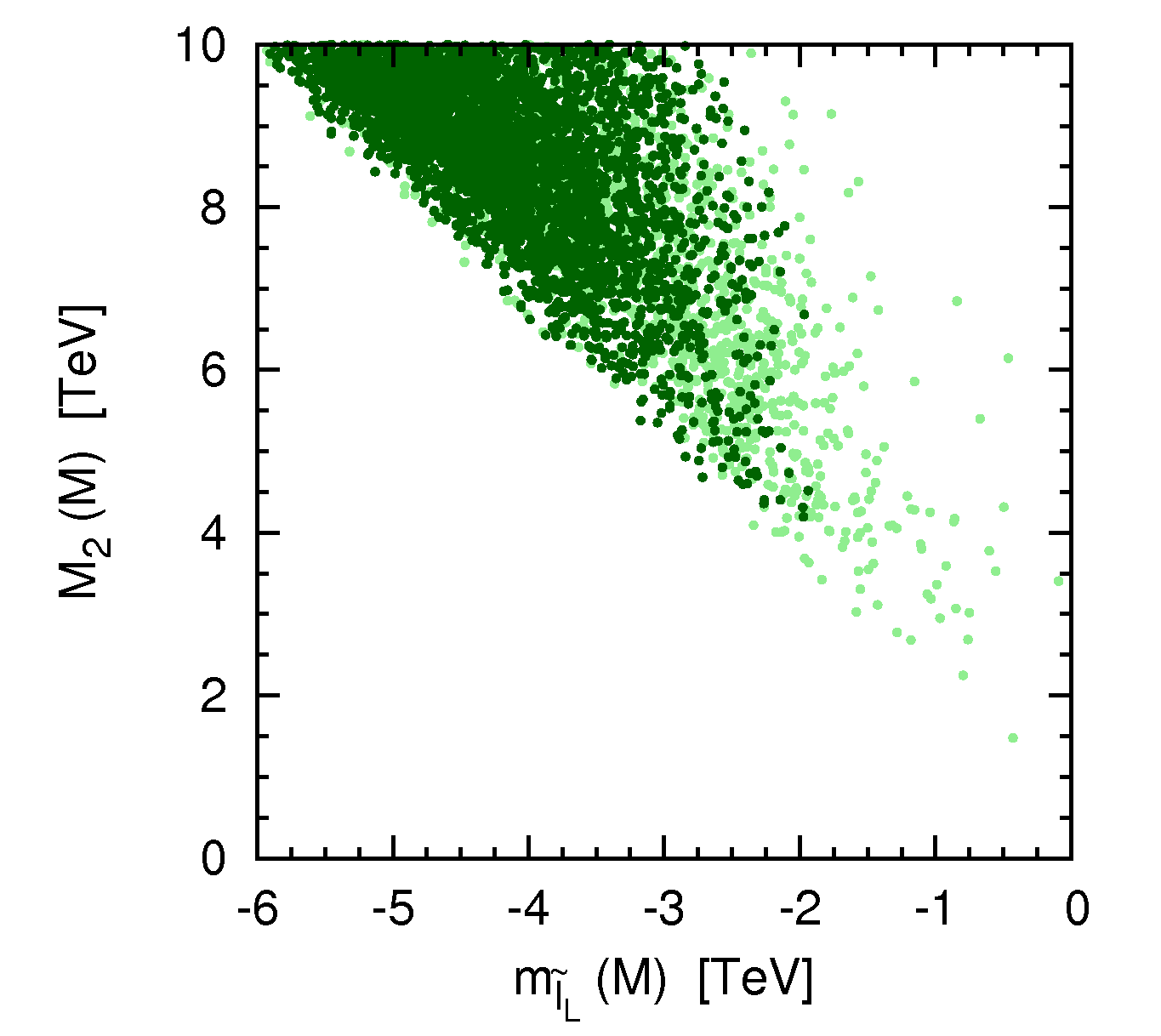}
\caption{\label{fig:GGM} Points with a selectron NLSP resulting from the scan of the GGM parameters
specified in \eqref{eq:ggm-scan}, displayed in terms of the messenger scale values of $m_{{\tilde \ell}_L}\equiv m^2_{{\tilde \ell}_L} /\sqrt{|m^2_{{\tilde \ell}_L}|}$ and $M_2$. The darker points satisfy, in addition, $m_h = 126\pm 3$ GeV.}  
\end{center}
\end{figure}

In Figure \ref{fig:GGM}, we show the result of a full numerical scan over the GGM parameter space in the plane defined by the high-energy values of
$m_{\widetilde{\ell}_L}\equiv m_{\widetilde{\ell}_L}^2 /\sqrt{|m_{\widetilde{\ell}_L}^2|}$ and $M_2$. 
The GGM parameters were  varied independently in the following ranges: 
\begin{align}
  (100~{\rm GeV})^2 \le |m^2_{\widetilde{\ell}_R}| \le (2~{\rm TeV})^2, \quad
 (1~{\rm TeV})^2 \le & |m^2_{\tilde Q}| \le (10~{\rm TeV})^2 , \quad
 0 \le |m^2_{\widetilde{\ell}_L}| \le (10~{\rm TeV})^2 , \nonumber \\
  100~{\rm GeV} \le M_1 \le 2~{\rm TeV}, \quad
 100~{\rm GeV} \le & M_2 \le 10~{\rm TeV}, \quad
 1~{\rm TeV} \le M_3 \le 10~{\rm TeV}, \quad \nonumber \\
  5 \le \tan\beta \le 50, & \quad 10^5 ~{\rm GeV} \le M \le 10^{15} ~{\rm GeV}.
\label{eq:ggm-scan}
\end{align}
Notice that both signs of $m^2_{\tilde f}$ were considered. 
In Figure \ref{fig:GGM}, the light-green points correspond 
to a selectron NLSP with $m_{\widetilde{\tau}_1}- m_{\widetilde{\ell}_R}  \ge 20$ GeV and
which fulfill the basic phenomenological requirements: no tachyons at the EW scale, successful EWSB etc. Moreover we discard all points which have superpartners heavier than $10\text{ TeV}$, thus imposing an indirect mild upper bound on the soft parameters at the messenger scale. The dark-green points, in addition, account 
for the observed Higgs mass, up to theoretically uncertainties:  
$m_h = 126\pm 3$ GeV. 

Our scan confirms that a selectron NLSP can be obtained as a consequence of large negative values of $m^2_{\widetilde{\ell}_L}$, while $M_2$ also has to be large in order to avoid tachyonic LH sleptons in the IR. In particular, 
we see from Figure \ref{fig:GGM} that the lowest possible value of $M_2$ that is compatible with a non-tachyonic spectrum, rapidly increases as $|m^2_{\widetilde{\ell}_L}|$ increases. 
The observed Higgs mass needs rather heavy stops
in GGM, and thereby a large $\mu$. Therefore we can obtain $m_{\widetilde{\tau}_1}> m_{\widetilde{\ell}_R}$
and $m_h = 126\pm 3$ GeV (dark points) only for a sizeable $|m^2_{\widetilde{\ell}_L}|$ (i.e.~large $\Delta_R$), 
as we expected from the results shown in Figure \ref{fig:mutb}. 

From this scan, we observed that 
the low-energy value of the selectron mass can be as light as $\approx 200$ GeV and 
a selectron NLSP in GGM models is only possible for $M \gtrsim 10^{7} ~{\rm GeV}$, i.e. for a sufficiently long RG running. This implies that the ${\tilde\ell}_R$ decay to lepton and gravitino is never prompt, as will be discussed in Section \ref{sec:collider}.

\subsection{Deflected Models}\label{sec:Deflection} 

We define ``deflected'' models as those models of GM that  feature 
additional contributions to the Higgs masses at the messenger scale, besides the GGM one of Eq.~\eqref{eq:higgses}. These additional contributions are due to the presence of extra superpotential interactions between the hidden sector and the Higgs sector.
In particular, the new interactions can generate $\mu$ and $B_{\mu}$ at the messenger scale, thus being good candidate to solve the $\mu$ problem in GM \cite{Dvali:1996cu,Csaki:2008sr,Komargodski:2008ax,DeSimone:2011va}. Another nice feature of these kind of models is the possibility of generating non-zero A-terms at the messenger scale \cite{Chacko:2001km,Evans:2010kd,Evans:2011bea,Evans:2012hg,Kang:2012ra,Craig:2012xp}. A complete study of the threshold corrections at the messenger scale that one can get in this class of models have been performed in \cite{Evans:2013kxa}.
Without entering the details of any specific model, we discuss here the general features of this setup, which are relevant for the selectron NLSP scenario. In  Section \ref{messmodel}, we will discuss simple messenger models that explicitly realize this spectrum.  

Equation \eqref{RGDelta} shows that the selectron NLSP scenario can be achieved by means of tachyonic boundary conditions for the down-type Higgs, $m^2_{H_{d}}<0$. For this reason we focus our attention on modifications of the high-energy thresholds of GGM only for the Higgs soft masses, assuming for the moment that all the other soft masses are not deflected. This simplifying assumption is not necessarily realized in concrete
models, as we will show in Section \ref{messmodel}.
We can account for the Higgs deflections by adding two independent parameters at the messenger scale: 
\begin{align}
m^2_{H_{u/d}}(M)=m^2_{\widetilde{\ell}_L} + \Delta^2_{u/d}\, .
\label{mHud}
\end{align}
Clearly, these new contributions invalidate the GGM relation (\ref{eq:higgses}). Moreover, they deform the hypercharge sum rule in the sfermion sector $\text{Tr}(Ym^2_{\widetilde{f}})=S(M)$ by introducing a non-zero Fayet-Ilioupoulos term at the messenger scale,   $S(M)=  \Delta^2_{u} - \Delta^2_{d}$.

If $\Delta_d^2$
is negative and sufficiently large, 
then the RHS of \eqref{RGDelta} can be negative even if both the RH and the LH slepton squared masses are positive, leading to selectron NLSP.
This is depicted in the left panel of Figure \ref{fig:LL-est}, where we plot both the leading-log estimate of Eq.~(\ref{eq:LL}) and the exact RG-solution for the splitting of the RH stau and selectron soft masses. 

However, having  $m^2_{H_{d}}$ tachyonic can affect the EWSB.
In the MSSM, the two EWSB conditions can be written as
\bea
&&\frac{m_Z^2}{2}=-|\mu|^2-\frac{m_{H_{u}}^2 \tan^2 \beta+m_{H_d}^2}{\tan^2 \beta-1}\, ,\label{muEWSB}\\
&& \frac{2 B_{\mu}}{m_{A}^2}=\frac{2 \tan \beta}{1+\tan^2 \beta} \qquad \text{with} \qquad m_A^2=2 |\mu|^2+m_{H_u}^2 + m_{H_d}^2\, ,\label{AEWSB}
\eea
where every parameter is evaluated at the EW scale. For large to moderate values of $\tan\beta$, the term in \eqref{muEWSB} containing the down-type Higgs soft mass can be neglected and one obtains the approximate expression
\be
|\mu|^2 \simeq -m_{H_{u}}^2\,.
\ee 
Inserting this relation into the mass formula for the CP-odd Higgs, one gets that, at the EW scale, in the large $\tan\beta$ limit
\be
m_{A}^2 \simeq  m_{H_d}^2-m_{H_u}^2\,, \label{Amass}
\ee
\begin{figure}[t]
\begin{center}
\includegraphics[width=0.48\textwidth]{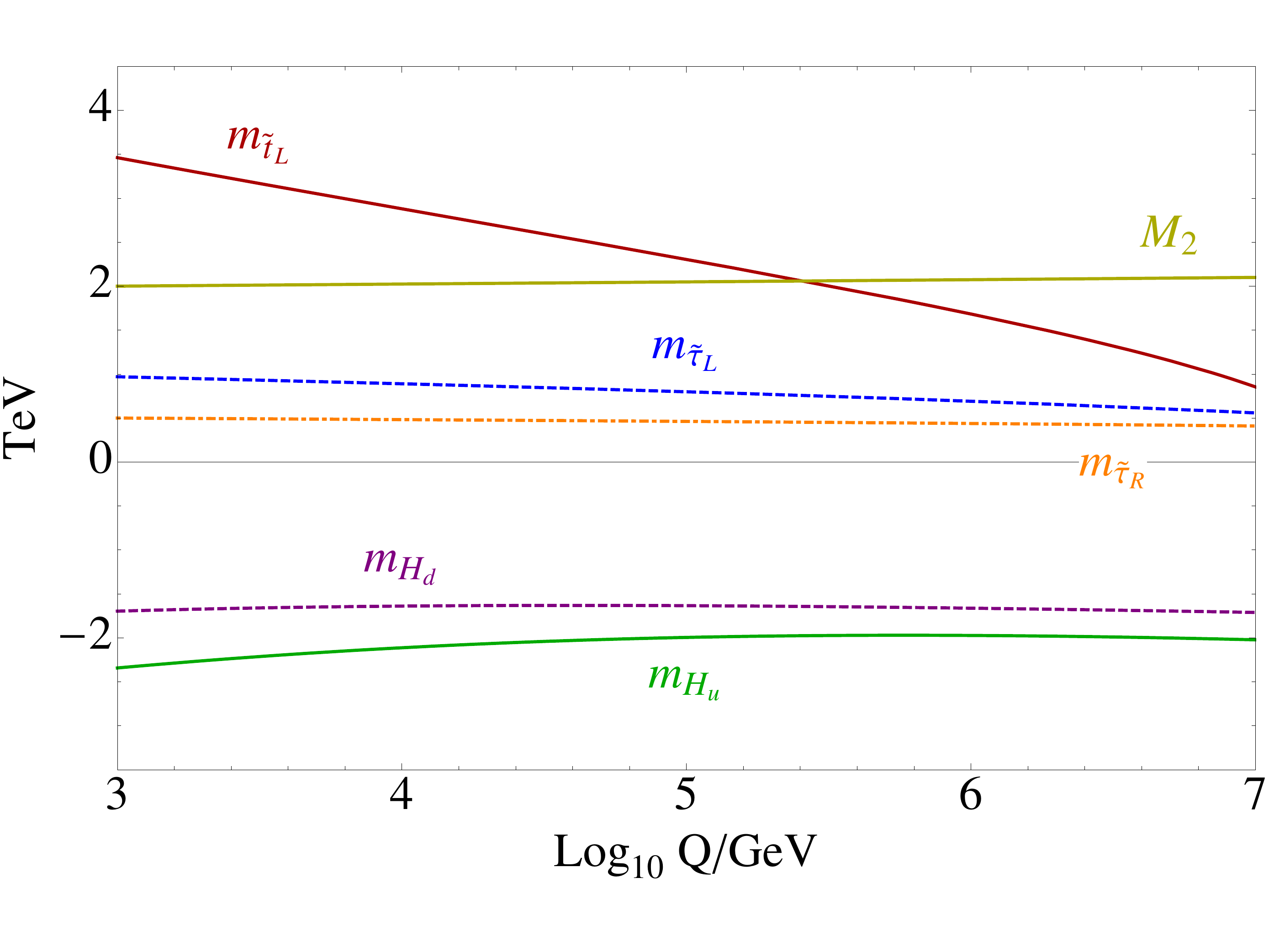}
\hspace{10pt}
\includegraphics[width=0.48\textwidth]{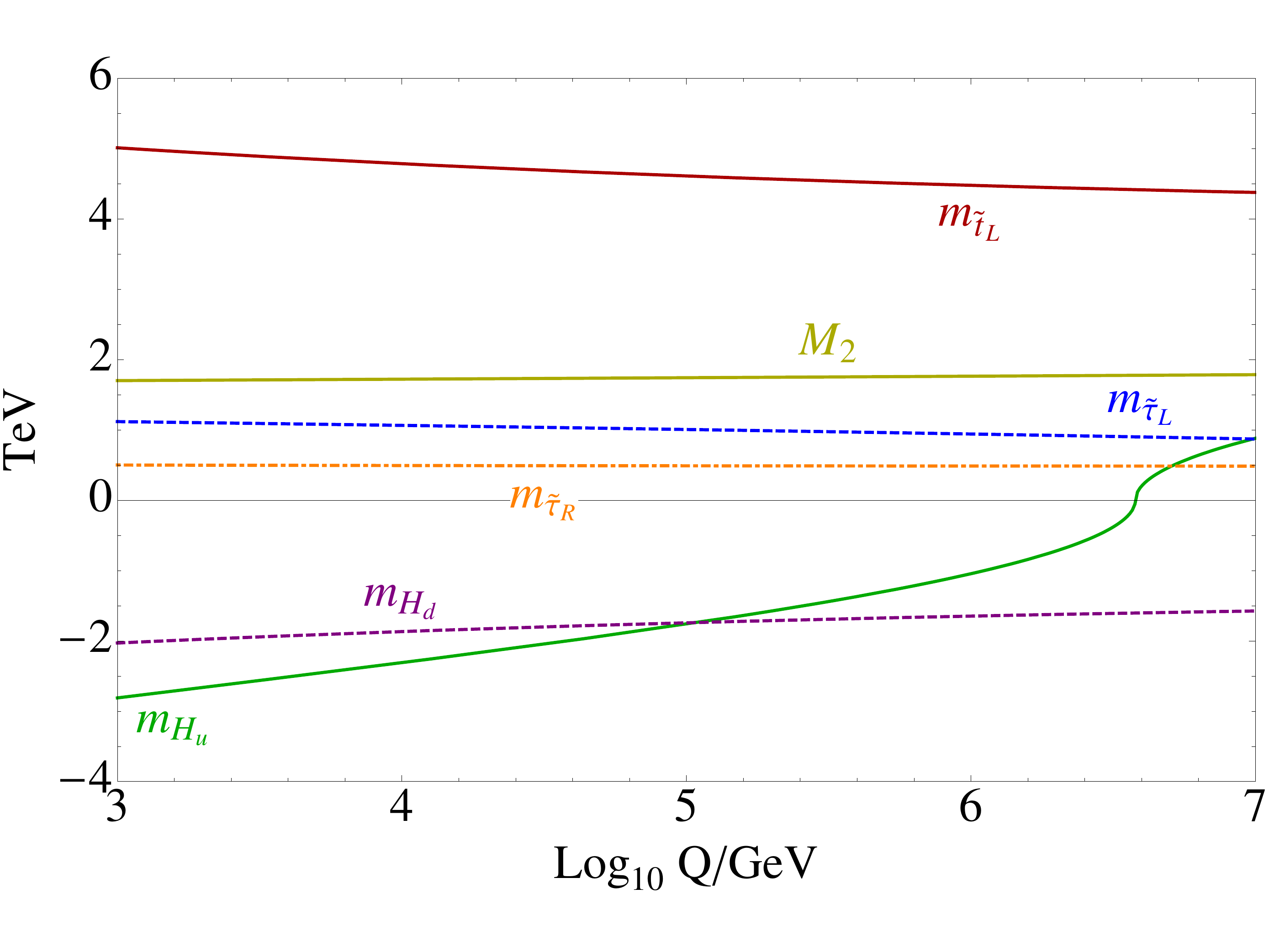}
\caption{Examples of RG flows from the messenger scale $M$ to $M_{S}=\sqrt{m_{\tilde{t}_L} m_{\tilde{t}_R}}$ for points with selectron NLSP in ``deflected'' scenarios. We show two examples with $M=10^{7}\text{ GeV}$ and $\tan\beta=30$. In both cases $\Delta_R=200\text{ GeV}$, $m_{\widetilde{\tau}_R}=500\text{ GeV}$ and $M_{1}=550 \text{ GeV}$.  \label{fig:RGflowsExample2}  } 
\end{center}
\end{figure}
indicating that $m_{H_d}^2<0$ can potentially lead to a tachyionic CP-odd Higgs.

As a first ``tree level" solution to this problem, 
equation \eqref{Amass} suggests that, in order to obtain $m_{A}^2>0$,\footnote{The current bound on $m_A$ from direct searches \cite{CMS-PAS-HIG-12-050} -- setting a more stringent constraint -- will be taken into account in the numerical analysis.} 
$m_{H_u}^2$ should also be negative at the messenger scale and, in absolute value, larger than $m_{H_d}^2$. This is indeed a viable case and it is displayed in the left panel of Figure \ref{fig:RGflowsExample2}. Note that, even if starting with a large and negative $m^2_{H_u}$, which induces a large $\mu$ at the EW scale, the left-right mixing in the stau mass matrix can always be suppressed by a large $m_{\widetilde{\ell}_L}$, which is not forced to start tachyonic, in contrast to the GGM case. 

Another possibility to circumvent the obstruction given by the requirement of a non-tachyonic $m_{A}$  is to enhance the negative contributions to $m_{H_u}$ (driven by terms $\propto y_t^2$) from the RG running, which are actually responsible for the radiative EWSB.
These radiative effects are summarized by the following RGEs:
\begin{eqnarray}
16\pi^2\frac{\text{d}}{\text{d}t} m_{H_u}^2&=&3 X_{t}-6 g_2^2|M_2|^2-\frac{6}{5}|M_1|^2+\frac{3}{5}g_{1}^2S\,, \\
16\pi^2\frac{\text{d}}{\text{d}t} m_{H_d}^2&=&3 X_{b}+X_\tau-6 g_2^2|M_2|^2-\frac{6}{5}|M_1|^2-\frac{3}{5}g_{1}^2S\,,
\end{eqnarray}
where the Fayet-Iliopoulos term $S$ is given in \eqref{FI} and where
\begin{equation}
X_{t/b}=2\vert y_{t/b}\vert^2\left(m^2_{H_{u/d}}+m^2_{\widetilde{t}_{L}/\widetilde{b}_{L}}+m^2_{\widetilde{t}_{R}/\widetilde{b}_{R}}+\vert A_{t/b}\vert^2\right)\,.
\end{equation}
Here we see that, beside the terms proportional to $S$, the contributions from the gauge interactions to the difference $m_{H_d}^2-m_{H_u}^2$ vanish. The RG equation for the difference in \eqref{Amass} is then given by
\begin{equation}
16\pi^2\frac{\text{d}}{\text{d}t}(m_{H_d}^2-m_{H_u}^2)=3(X_b-X_{t})+X_{\tau}-\frac{6}{5}g_{1}^2S\, .\label{RGCPodd}
\end{equation}
Hence, in models with $m_{H_d}^2$ negative, the problem of having $m^2_A <0$ at the EW scale can be alleviated for instance by
heavy stops, a large $A_t$ or a large gluino mass, which drive up the stop masses at low energies.
Interestingly, within the MSSM, the very same conditions 
are required in order to accommodate a SM-like Higgs scalar with mass around 126 GeV. 
In the right panel of Figure \ref{fig:RGflowsExample2} we display a possible solution with heavy stops in which the entire soft spectrum at the messenger scale is non-tachyonic, except for $m^2_{H_d}$, which is responsible for triggering the desired effect. This benchmark realizes, in a concrete setup, the spectrum of Figure \ref{fig:LL-est} (left).  

\begin{figure}[t]
\begin{center}
\includegraphics[width=0.48\textwidth]{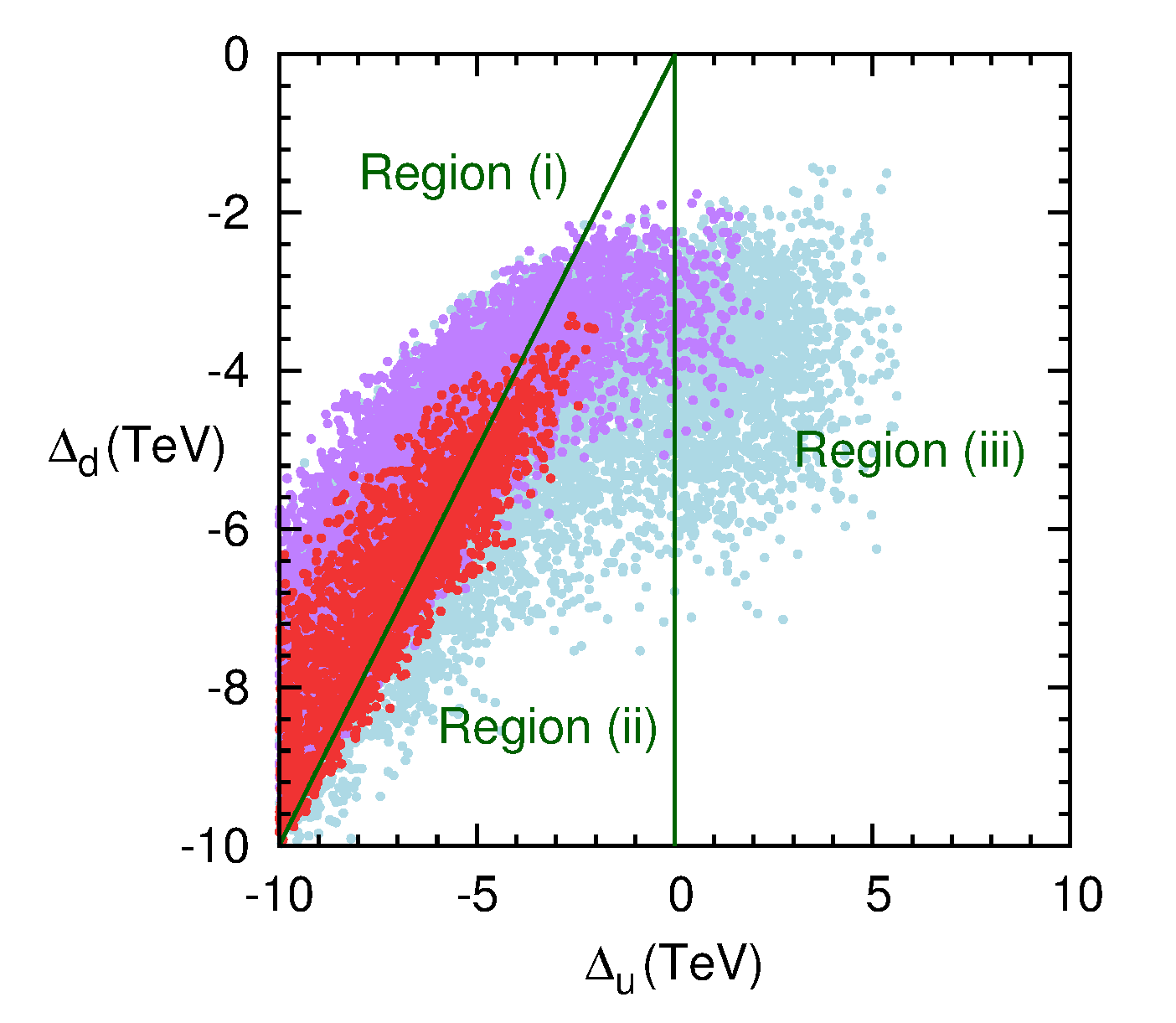}
\includegraphics[width=0.48\textwidth]{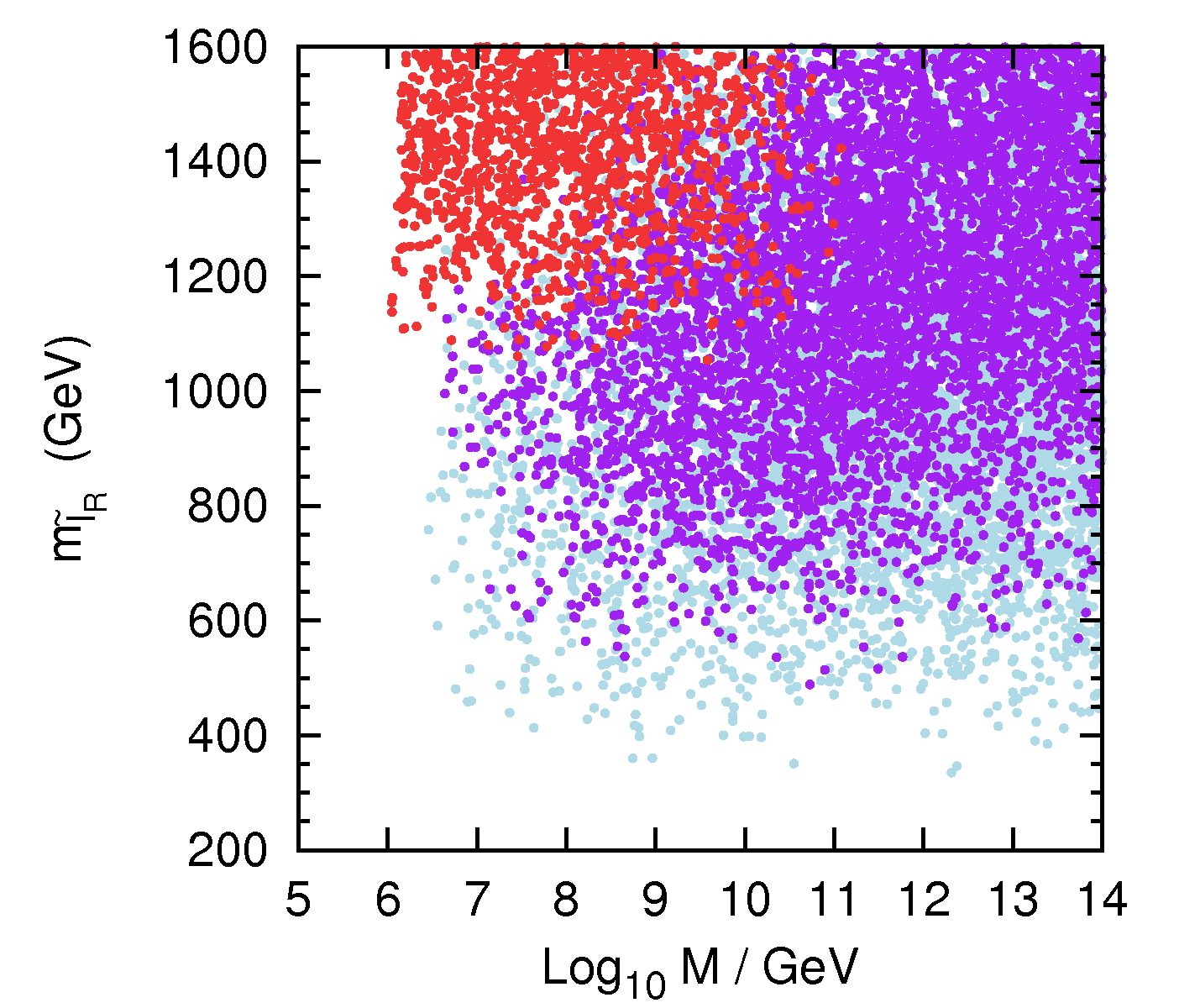}
\caption{\label{fig:semiplane}
Points with a selectron NLSP for the deflected model and for different choices for the high-energy parameters: 
$\Lambda_{G_{i}}=\Lambda_{S_{i}}\equiv \Lambda$ (red points), 
independent  $\Lambda_{G}$ and $\Lambda_{S}$ (purple points),
independent $\Lambda_{G}$, $\Lambda_{S_{3}}$ and $\Lambda_{S_{1,2}}$ (blue points).
The results are displyed in the $(\Delta_u,~\Delta_d)$ plane (left panel) and
in the $(M,~m_{{\widetilde \ell}_R})$ plane (right panel). See the text for further details.
}  
\end{center}
\end{figure}

Following the above discussion, we have performed a  scan of the UV parameters, again requiring a 
sizeable mass-splitting between the lightest stau and the selectron NLSP, $\ge 20$ GeV, as well as the other constraints. 
The points in the scan that exhibit a selectron NLSP are shown in Figure \ref{fig:semiplane}. In the left panel we display the plane $(\Delta_u,~\Delta_d)$, defined by:
\be
\Delta_{u/d}\equiv \frac{\Delta^2_{u/d} }{\sqrt{\vert\Delta^2_{u/d}\vert}}\,.
\ee
As is highlighted in the figure, three different viable regions with selectron NLSP can be identified, which correspond to different realizations of the EWSB:

\begin{itemize}
\item Region (i), $\Delta^2_u<0$ with $\vert \Delta^2_u\vert\geq\vert \Delta^2_d\vert$: this corresponds to the simplest (``tree-level'') solution to avoid $m^2_A<0$ at the EW scale, in which we allow for tachyonic up-type Higgs masses, at the messenger scale, which are larger in modulus than the down-type one, cf.~the left panel of Figure \ref{fig:RGflowsExample2}. 
\item Region (ii), $\Delta^2_u<0$ with $\vert \Delta^2_u\vert<\vert \Delta^2_d\vert$: the negative contribution to the up-type Higgs mass at the messenger scale is lower in modulus than the one to the down-type Higgs, so that we access an intermediate region in which the problem of the tachyonic mass for the CP-odd Higgs is solved partially by radiative contributions that can be induced by large stops or a large $A_{t}$.  
\item Region (iii), $\Delta^2_u\geq0$: in this case, a non-tachyonic value of the CP-odd Higgs mass, at the EW scale, is realized purely by radiative contributions coming from large stop masses (cf.~right panel of Figure \ref{fig:RGflowsExample2}) or a large $A_{t}$. 
\end{itemize}

In the right panel of Figure \ref{fig:semiplane}, the result of the scan is shown in terms of the messenger scale $M$ and the low-energy 
value of the selectron NLSP mass. 
Important hints on the model building requirements can be extracted by identifying the parameters in Equations  \eqref{gmass} and \eqref{sfmass} which have to be independent at the messenger scale in order to realize each of the above regions. 
Points with different colors in Figure \ref{fig:semiplane} correspond to different choices: 
the red points correspond to the simplest models of GM with $\Lambda_{G_{i}}=\Lambda_{S_{i}}\equiv \Lambda$, 
the purple points to a two-scale setup with two separate parameters $\Lambda_{G}$ and $\Lambda_{S}$ controlling the gaugino and sfermion masses, respectively, and the light-blue points correspond to the case where sfermion mass unification is relaxed by taking $\Lambda_{S_{3}}\neq\Lambda_{S_{1,2}}$.
The SUSY breaking parameters were varied in the range $10^4\div10^6$ GeV for all the three cases.
For the other parameters we took: $5\le \tan\beta \le 50$, 
$2\times{\rm max}(\Lambda_{G_i},\Lambda_{S_i}) \le M \le 10^{15}$ GeV.

As we can see from Figure \ref{fig:semiplane}, the selectron NLSP scenario in region (i) can be obtained even with $\Lambda_{G_{i}}=\Lambda_{S_{i}}=\Lambda$ (red points), 
while region (ii) only marginally and region (iii) is not accessible in this case.
In fact, effective radiative corrections are very much constrained by the fact that there is only one scale that controls the whole soft spectrum. A large $\Lambda$ would be needed in order to obtain a non-tachyonic CP-odd Higgs with $\Delta^2_u>0$, but that would also increase the universal contribution to the slepton masses, washing out the effects of the Yukawa interactions that might give a selectron NLSP.
Moreover, this scenario shares the phenomenological problems of minimal GM, 
in particular the requirement of 
$m^2_S \equiv m_{\tilde{t}_1} m_{\tilde{t}_2} \simeq (5~{\rm TeV})^2$ needed in order to have $m_h \approx 126$ GeV. 
As is shown by the right panel of Figure \ref{fig:semiplane}, this translates into a lower bound on the scalar masses, in particular on the slepton mass ($m_{\widetilde{\ell}_R}\gtrsim 1$ TeV), again because the spectrum is essentially controlled by a single parameter. As a consequence, testing such a scenario at the LHC would be very challenging.

In order to access the region (iii), large radiative corrections to $m^2_{H_u}$ are necessary. 
The simplest possibility is to rely on the gaugino-driven contribution of the running, through a quite heavy gluino. This is possible by splitting $\Lambda_{G}$ and $\Lambda_{S}$, as is shown in Figure \ref{fig:semiplane} (purple points). 
Notice that this scenario realizes automatically gaugino and sfermion mass unification at the messenger scale and hence, it can be easily embedded in messenger models with a complete GUT structure. A common mass parameter for the gauginos however
implies a rather heavy spectrum, in particular $m_{\widetilde{\ell}_R}\gtrsim 500$ GeV, cf.~the right panel.

In order to realize region (iii) with lighter sleptons, we need further contributions from large stop masses and/or large $A_{t}$. 
This latter possibility will be discussed in an explicit model in the next section, since the extra interactions which generate large A-terms will also typically contribute to the sfermion masses, resulting in a sizeable deflection of the spectrum from the usual GM one. 
Here we consider only the possibility of splitting the colored sector so that heavy squarks can be obtained, while keeping the sleptons light. 
This can be done in two ways: 
either by relaxing the hypothesis of sfermion mass unification, i.e.~$\Lambda_{S_{3}}\neq\Lambda_{S_{1,2}}$, 
or by dropping gaugino mass unification, i.e.~$\Lambda_{G_{3}}\neq\Lambda_{G_{1,2}}$. For illustrative purposes, 
in Figure \ref{fig:semiplane}, we adopted the first possibility (light-blue points). As we can see, region (iii) can now be easily accessed. Moreover, the light-blue points correspond to slepton masses down to $m_{\widetilde{\ell}_R}\gtrsim 300$ GeV. 

As a final remark, Figure \ref{fig:semiplane} shows that, within these models it is rather difficult  to obtain our effect
for $M\lesssim 10^7$ GeV, especially for light sleptons. Therefore, like in the case of GGM, these models typically 
predict the NLSP decay to be displaced from the interaction point (either inside or outside of the detector), 
as will be clear from the discussion in Section \ref{sec:collider}. Note, however, that the lower bound on the messenger scale can in principle be circumvented if we allow the soft masses for the scalars, other than the Higgses, to be tachyonic at the messenger scale.

\section{Realizations in terms of messengers models}\label{messmodel}
In this section we investigate possible concrete realizations of the selectron NLSP scenario in terms
of weakly coupled messenger models, possibly directly 
coupled to the two Higgs doublets of the MSSM.

As mentioned in the previous section, it is possible to obtain selectron NLSP
with standard (non-tachyonic) UV boundary conditions for the soft masses at the price of accepting a large tuning of the UV parameters.
The only requirements consist of long running, i.e.~a large messenger scale, and large $\tan\beta$.
From a model building perspective, 
this case corresponds to usual models of (general) 
gauge mediation, and we will not discuss it any further here.

The other two cases we have described rely on negative squared masses in the UV,
and are in some sense complementary.
In the first case, which is within the definition of GGM, the squared masses of the left-handed
sleptons are negative in the UV, and 
equal to the down-type Higgs squared mass; we will discuss this in Subsection \ref{GGMmodel}.
In the second case,  we generate negative squared masses only for the
down-type Higgs by coupling the Higgses to some hidden sector fields,
thus going beyond the pure GGM paradigm; we will explore this option in Subsection \ref{deflectedmodel}.
 
 Finally, in Subsection \ref{sec:Yanagida}, we study a model which also includes extra contribution
 to the A-terms. This represents 
 the 
most economical model 
that features promptly decaying selectron/smuon co-NLSP, a correct Higgs mass and also relatively light stops.

\subsection{Boundary conditions with tachyonic slepton masses}
\label{GGMmodel}
The SUSY breaking parameters $\Lambda_{G_i}$ and $\Lambda_{S_i}$ 
determine the UV pattern of soft masses of GGM, and here we investigate whether it is possible
to obtain the desired UV boundary conditions in models with weakly coupled messengers.
The purpose of this section is to provide a proof of existence, 
without the ambition of being complete.

As was explained in Section \ref{sec:GGM},
in order to obtain selectron NLSP
it is sufficient to 
consider UV tachyonic boundary conditions for the left-handed sleptons.
There are several mechanisms able to generate a negative squared mass for the
scalars in GM.
One possibility consists of considering gauge messengers,
as explained in \cite{Intriligator:2010be,Buican:2009vv}. This would require to specify
the embedding of the SM gauge group into the unification group,
as well as the mechanism that breaks it.
Another option is to consider models where the Supertrace on the messengers is 
non-vanishing and positive. 
This would induce a negative contribution to the scalar
soft masses in the MSSM, as for instance in models of direct gauge mediation 
\cite{ArkaniHamed:1997jv}.
However, as shown in \cite{Poppitz:1996xw}, in minimal messenger models 
this contribution is divergent and it introduces logarithmic dependence on the UV cut-off $\Lambda$
\begin{equation}
\delta  m^2_{\widetilde{f}}(M)= -2\sum_{i=1}^{3}C_{\widetilde{f}_{i}}\frac{g^4_{i}(M)}{(4\pi)^4}\text{Str} \mathcal{M}_\text{Mess} \log\Lambda\ .
\end{equation} 

A possible way to soften the logarithmically divergent
contribution is to UV complete the theory, for instance with models
of semi-direct gauge mediation, where the messengers couple to 
the supersymmetry breaking sector only through another extra gauge group \cite{Seiberg:2008qj,Argurio:2009ge}.
However these models are plagued by the gaugino screening problem \cite{ArkaniHamed:1998kj,Argurio:2009ge},
and hence are not useful in our setup.

Finally, even if the Supertrace on the messenger sector is vanishing, 
the simultaneous D and F term breaking of at least two pairs
of messenger fields can lead to negative squared masses for the
sfermions \cite{Buican:2008ws}.
In particular,
this requires the pairs of messengers to be oppositely charged under an extra 
gauge group, with non-vanishing D-term breaking.
This is the strategy we adopt in the following.\footnote{It is interesting to note that the weakly coupled possibilities discussed here
are all characterized by the presence of heavy massive vector bosons in the hidden sector.}

We also demand gauge coupling unification
to be preserved. This would require the messengers
to belong to complete
representations of the unification group or to some ``magic'' set, as discussed in \cite{Martin:1995wb,Calibbi:2009cp}.

We propose the set of messengers reported in Table \ref{messengers},
which are vectorlike pairs in the fundamental of the SM gauge groups.
 \begin{table}
 \begin{center}
\begin{tabular}{|c|c|c|c|c|c|}
\hline
\# of pairs   & $SU(3)$ & $SU(2)$ & $U(1)_Y$ & $U(1)_H$ \\
\hline 
$3$ & $\mathbf{3} + \mathbf{\bar 3}$ & $\mathbf{1}$ & 0  & 0 \\
\hline
$1$ & $\mathbf{1}$ & $\mathbf{2} + \mathbf{\bar 2}$  & 0  & $ 1$ \\
\hline
$1$ & $\mathbf{1}$ & $\mathbf{2} + \mathbf{\bar 2}$  & 0  & $- 1$ \\
\hline
$1$ & $\mathbf{1}$ & $\mathbf{2} + \mathbf{\bar 2}$  & 0  & $0$ \\
\hline
$10$ & $\mathbf{1}$ & $\mathbf{1}$ & $\pm \frac{1}{2}$  & 0 \\
 \hline  
\end{tabular}
\end{center}
\caption{\label{messengers}
The set of weakly coupled messengers.}
\end{table}
One can easily show that this set of fields 
induces the same shift in the beta function coefficients of the $SU(3)$, $SU(2)$ and $U(1)_Y$ gauge couplings.
As a consequence, unification at the usual MSSM GUT scale is preserved, 
even though the messengers do not form a complete GUT representation.
In the following we will consider slightly different mass scales for some messengers,
assuming that the consequent thresholds induced on the running of the gauge couplings are negligible.

As is shown in Table \ref{messengers},
two out of the three $SU(2)$ charged messengers are also charged 
(with opposite charge) under an extra $U(1)_H$ gauge group, with a non-vanishing D-term.
Moreover,
we assume the following superpotential couplings of the messenger fields to 
some spurions, i.e.~the $X$'s:
\be
W= X_1 \sum_{j=1}^{10} \tilde \Phi_{1}^j \Phi_{1}^j 
+X_2^+ \tilde \Phi_{2}^+ \Phi_{2}^+
+
X_2^- \tilde \Phi_{2}^- \Phi_{2}^- 
+
X_2^0 \tilde \Phi_{2}^0 \Phi_{2}^0 +
X_3 \sum_{i=1}^3 \tilde \Phi_{3}^i \Phi_{3}^i \,,
 \ee
 where the subscripts refer to the gauge group under which the messenger field is charged and
 the superscripts of the $SU(2)$-charged fields indicate
 their charge under the $U(1)_H$ gauge group.
 The spurions take the following form,
 \be
X_3= M+\theta^2 F_3 
\, ,
X_2^0= M+\theta^2 F_2^0
\, ,
X_2^+= M+\theta^2 F_2^+ 
\, ,
X_2^-= M'+\theta^2 F_2^-
\, ,
 X_1= M+\theta^2 F_1\,,
 \ee
 with the choice $M' > M$ such that the contribution to the soft masses will be negative.
 This configuration leads to the following SUSY-breaking parameters, which determine the soft terms:
 \bea
&&\Lambda_{G_3}=3 \frac{F_3}{M}\,, \qquad \Lambda_{G_2}=\frac{F_2^0}{M}+\frac{F_2^+}{M}+ \frac{F_2^-}{M'}\,, \qquad 
\Lambda_{G_1}=3 \frac{F_1}{M}\,, \\
&&\Lambda_{S_3}^2= 3  \frac{F_3^2}{M^2}\,, \qquad 
\Lambda_{S_2}^2=\frac{(F_2^0)^2}{M^2} +\frac{(F_2^+)^2}{M^2}+ \frac{(F_2^-)^2}{(M')^2}+2  D_H \log \frac{M^2}{M'^2}\,,  \qquad \Lambda_{S_1}^2=3 \frac{F_1^2}{M^2}\,.
\nonumber
 \eea
 From these expressions it is clear that $\Lambda_{S_2}^2$ can be made negative
 if $D_H$ is sufficiently large. In order to avoid a large tuning, 
 we expect that in such situations, $|\Lambda_{S_2}|$ and $\Lambda_{G_2}$
 are of the same order. This is compatible with Figure \ref{fig:GGM}, 
 where $M_2 \gtrsim 2 |m_{\widetilde{\ell}_L}|$ in the region of selectron NLSP.
Depending on the value of $F_3$, the left-handed squarks can have positive or negative UV squared masses,
and we have seen in Section \ref{sec:GGM} that both cases are possible.

Finally, $F_1$ characterizes the Bino and the right-handed slepton masses.
The hierarchy between the Bino and the right-handed sleptons masses
is determined by the length of the RG flow, i.e. by the messenger mass $M$.
Since the effective number of messengers
in the $U(1)_Y$ sector is $3$, we can estimate that the Bino will be heavier
than the selectron/smuon as long as $M \leq 10^{10}$ GeV.\footnote{If needed, one can introduce a D-term breaking also for the $U(1)_Y$ messengers in order to reduce
the contribution to the right-handed sleptons and make the Bino/slepton mass ratio bigger.}

In order to verify that the model presented here can realize selectron NLSP, 
we performed a numerical scan by fixing $\Lambda_{G_3}=10^6$ GeV
and varying the other SUSY breaking parameters and the messenger mass.
We indeed find selectron NLSP in the expected region, i.e. for $\Lambda_{S_2}^2$ large and negative.
We do not present the results here since the qualitative features are very similar to the ones discussed 
after Figure \ref{fig:GGM} in Section \ref{sec:GGM}, once translated in terms of the UV soft masses.

\subsection{Tachyonic down-type Higgs mass}
\label{deflectedmodel}

Parametrizing the extra contributions to the two Higgs doublets of the MSSM, i.e.~in addition to the usual GM contributions, as $\Delta_u$
and $\Delta_d$, in Figure \ref{fig:semiplane},
we already identified the three possible interesting regions, characterized by different values
of $\Delta_u$ compared to a negative and typically large $\Delta_d$. In what follows, we survey the possibilities to induce a negative and large 
$\Delta_d^2$ at tree level, or at loop level, in models of weakly coupled messengers coupled to the Higgs sector.
For every scenario we comment how these extra couplings in the Higgs sector affect the other 
dimensionful parameters
characterizing the Higgs potential and the
sparticle soft masses.

\paragraph{Tree level $\Delta_d^2$}
Tree level contributions to the Higgs soft masses can be obtained by mixing the Higgs fields with
messengers coupled to a SUSY-breaking spurion.
A generic superpotential realizing this possibility is the one considered e.g.~in \cite{Komargodski:2008ax,Evans:2011bea}:
\be
\label{tree_level}
W= X \Phi_u \tilde \Phi_d+ \mu_u \tilde \Phi_d H_u+\mu_d \Phi_u H_d + \mu H_u H_d +W_{\rm Yukawa}\,,
\ee
where $X=M+\theta^2 F$ is a spurion superfield, and we assume a canonical Kahler potential for 
the Higgses and the messengers.
In the limit of large $M$ we can integrate out the messengers, resulting in 
\bea
W&=&W_{\rm MSSM}-\frac{\mu_u \mu_d}{X} H_u H_d\,,\\
K&=& K_{can}(H_u,H_d)+\left(\frac{\mu_u^2}{X X^{\dagger}}\right) H_u H_u^{\dagger} +\left(\frac{\mu_d^2}{X X^{\dagger}}\right) H_d H_d^{\dagger}\,,
\eea
where we have assumed real $\mu_d$ and $\mu_u$.
This leads to the following soft terms, at leading order in $F$,
\be
\delta \mu=\frac{\mu_d \mu_u}{M}\,, \qquad B_{\mu}=\frac{\mu_u \mu_d F}{M^2}\,, \qquad 
\Delta_{u,d}^2=-\mu_{u,d}^2\frac{F^2}{M^4}\,,
\qquad
a_{u,d}=\frac{\mu_{u,d}^2 F}{M^3} \,,
\ee
where $a_{u,d}$ denote the coefficient of the term $F_{u,d}^{\dagger} H_{u,d}$ in the Lagrangian,
giving rise to $A_b, A_{\tau}$ once we integrate out $F_d$. The A-terms
are nevertheless suppressed by extra powers of ${\mu_{u,d}}/{M}$ with respect to the soft masses.
Given the induced negative mass terms for both Higgs doublets, this model 
covers region (i) and (ii) of Figure \ref{fig:semiplane}. However, it also generates non-negligible contributions
to $\mu$ and $B_{\mu}$. 
Hence, the EWSB condition puts some constraint on
these parameters, which could possibly be circumvented by adding another sector to provide the appropriate contributions to $\mu$
and $B_{\mu}$ only. 

A more economical realization, which 
suppresses these extra contributions,
 can be achieved by 
imposing an R symmetry 
(broken by the VEV of the spurion $X$)
with charges such that the $\mu_u$ term of the superpotential is forbidden:\\
\be
\begin{tabular}{|c|c|c|c|c|c|}
\hline
    & $\Phi_u$ & $\tilde \Phi_d$ & $H_d$ & $H_u$ & $X$ \\
\hline 
 $U(1)_R$ & -1 & +1 & 1 & 1 & 2 \\
 \hline  
\end{tabular}
\ee\\
This assignment can realize models of region (ii), with large and negative $\Delta_d^2$,
vanishing $\Delta_u^2$, and with vanishing $B_{\mu}$ at the messenger scale.

Notice that the superpotential in (\ref{tree_level}) also induces soft masses for EW gauginos and sfermions 
with the usual GM formulas, since $\Phi_u$ and $\tilde \Phi_d$
are charged under $SU(2) \times U(1)$. In the limit $\mu_d \ll M$ these contributions
are the ones of a minimal GMSB model with $\Lambda_{1,2}={F}/{M}$.

Since the messengers have the same quantum numbers as 
the Higgs doublets, we could add extra Yukawa couplings
between the messengers and the matter superfields. We will explore this option in the explicit example of Section \ref{sec:Yanagida}.


\paragraph{$\Delta_d^2$ at loop level}
\label{looplevel}
In order to realize our selectron NLSP scenario, we have seen that the
negative squared mass for the down-type Higgs should generically be sizable, see Figures \ref{fig:LL-est}
and \ref{fig:semiplane}.
This implies that if we aim at obtaining this term from quantum corrections, we typically
need a sizeable SUSY breaking scale $F$. Quantum corrections will 
generate such terms if there is some messenger field, coupled with the Higgs superfields,
that acquires split masses when SUSY is broken, or,
in other words, which couples directly to the spurion. We would like to realize a scenario with $\Delta_d^2<0$ through a modular
structure, that can be attached to a given GM model, without affecting the rest of the soft spectrum.
This suggests to
focus on models with only singlet chiral superfields coupled to a spurion, so that the other soft masses are not modified by the GM effects.
The simplest superpotential that can achieve this is:\footnote{We could have envisaged more
complicated models with other doublets and singlets. Since this simple example already gives rise to negative
soft masses for the Higgses, we chose to restrict to the most economical possibility.}
\be
\label{1loopmodel}
W=  X S_2^2 + m S_2 S_1 + \lambda S_1 H_u H_d + W_{\rm MSSM}\,.
\ee
The trilinear coupling of the Higgses resembles the one in the NMSSM. 
However, here the fields $S_1$ and $S_2$ do not acquire any expectation value,
so no $\mu$ term is generated, it originates from a separate sector.
The SUSY breaking spurion $X=M + \theta^2 F$ induces one loop soft masses
for the Higgses.
Note that there is an $R$ symmetry, broken by the $X$ VEV, such that $R(H_u)+R(H_d)=2$
and $R(X)\neq0$, implying that $B_{\mu}$ cannot be generated 
at leading order \cite{Komargodski:2008ax}.
We can compute the one loop Kahler potential for this model, after integrating out the field $S_1$
and $S_2$,
\be
\label{Keff1loop}
K_{1-loop} = -\frac{1}{32 \pi^2} \Tr \mathcal{M} \mathcal{M}^{\dagger} \log \frac{\mathcal{M} \mathcal{M}^{\dagger} }{\Lambda^2}\,,
\ee
where $\mathcal{M}$ is the SUSY mass matrix.
From this we can extract the wave function renormalization for the Higgs superfields, as a
function of the spurion by expanding 
the effective Kahler potential up to the quadratic order in the Higgses:
\be
K_{eff}= \mathcal{Z}_u H_u H_u^{\dagger}+\mathcal{Z}_d H_d H_d^{\dagger}+(\mathcal{Z}_{ud} H_u H_d+{\rm h.c.})\,.
\ee
The relevant part of the Higgs wave function renormalizations is
\be
\small{
\mathcal{Z}_{u,d}
=\frac{\lambda^2}{32 \pi^2}
 \left(
 \sqrt{\frac{X X^{\dagger}}{ \left(m^2+X X^{\dagger} \right)}} 
 \log \left[\frac{m^2+2 X X^{\dagger}+2 \sqrt{X X^{\dagger} \left(m^2+X X^{\dagger}\right)}}{m^2+2 X X^{\dagger}-2 \sqrt{X X^{\dagger} \left(m^2+X X^{\dagger}\right)}}\right]\right)
}
\ee
and $\mathcal{Z}_{ud}=0$.
Expanding these expressions in $\theta^2$ and $\bar \theta^2$
one can extract the soft terms.
Here we show the result in the limit $M \ll m$.
As expected, $\mu$ and $B_{\mu}$ are vanishing, while the
Higgs soft masses and the A-terms, at first order in the SUSY breaking scale and in $M/m$, are 
\be
\label{tachy1loop}
m^2_{H_d}= m^2_{H_u}= -\frac{\lambda^2}{8 \pi^2} \frac{F^2}{m^2}\left(1+O\left(\frac{M^2}{m^2}\right) \right)\,,
\quad 
a_{u,d}=\frac{\lambda^2}{8 \pi^2} \frac{F M}{m^2}\left(1+O\left(\frac{M^2}{m^2}\right) \right)\,.
\ee
The soft masses are negative, and the $A$-terms are suppressed by $M/m$.
Note that in the opposite regime, i.e. $M \gg m$, the soft masses would instead be positive.

The contributions (\ref{tachy1loop}) generate negative and equal $\Delta_u$ and $\Delta_d$
at one loop, realizing the scenario at the boundary between region (i) and (ii) in Figure \ref{fig:semiplane}.
Note that this is the main modification to the sparticle spectrum induced by the extra
couplings and superfields, as the latter ones are singlets.
In this sense the superpotential (\ref{1loopmodel}) can be considered as a module 
to add to usual GGM scenarios, in order to obtain a selectron NLSP.

\subsection{The simplest example: Yukawa-deflected gauge mediation}\label{sec:Yanagida}
We discuss here a simple example, based on the previous observations, that generically leads
to promptly decaying selectron/smuon co-NLSP. The model also accommodates the observed Higgs mass
and features a light stop, 
as extensively discussed in recent literature \cite{Evans:2011bea,Evans:2012hg,Kang:2012ra,Evans:2013kxa,Abdullah:2012tq,Calibbi:2013mka}.

We consider the following superpotential for the Higgses and the messengers,
\be
\label{summW}
W=X \Phi_u \tilde \Phi_d+\mu' \Phi_u H_d + \mu H_u H_d + \lambda_t \Phi_u Q_3 \bar u_3  +W_{\rm Yukawa}
\ee
where as usual $X$ is a spurion $X=M+\theta^2 F$.
Beside the mixing terms among the Higgses and the messengers, 
we consider only one extra top-like matter-messenger coupling,
following  \cite{Evans:2011bea,Evans:2012hg}.

Let us remark that, unlike the models considered so far, 
the presence of matter-messenger couplings generates new contributions to the 
sfermion mass matrices whose flavour structure can give rise to effects beyond minimal flavour violation, 
in principle spoiling the usual flavour protection of GM. 
However, the new flavour effects are completely under control
as far as the new couplings feature a hierarchical structure that resembles the ordinary
Yukawa matrices \cite{Abdullah:2012tq,Calibbi:2013mka}, as is the case here where we consider
a single $\mathcal{O}(1)$ top-like coupling $\lambda_t$.
The absence of other Yukawa-like couplings can be enforced
with a global $U(1)$ under which the messengers and the Higgses are
appropriately charged (see e.g. \cite{Evans:2011bea, Calibbi:2013mka}).

As already discussed,
the mixing term $\mu'$ induces the negative deflection for $m^2_{H_d}$ at tree-level,
\be
(\Delta^2_d)_{\rm tree} = 
-\mu'^2 \frac{F^2}{M^4}\left(1+\mathcal{O}({F^4}/{M^8})\right)~.
\label{eq:tree}
\ee
Note that this deflection is suppressed in the regime $F/M^2 \ll 1$, i.e.~for moderate to heavy mediation scales, unless we assume $\mu' \gg \mu$.
Hence, we argue that we will need small to moderate $M$ in order
for this effect to be sizable and lead to a selectron NLSP.

The superpotential (\ref{summW}) induces also other soft terms, which we now review.
At one loop, the coupling $\lambda_t$ 
generates top and bottom A-terms,
and also negative contributions to the stop masses, 
which are also suppressed for $F/M^2 \ll 1$ \cite{Evans:2011bea}:
\begin{align}
A_t = -\frac{3y_t}{16 \pi^2} |\lambda_t|^2  \frac{F}{M},~~&   A_b = -\frac{y_b}{16 \pi^2} |\lambda_t|^2 \frac{F}{M} \nonumber\\
(\Delta m^2_{{\tilde t}_L})_{\rm 1-loop}= -\frac{|\lambda_t|^2}{96 \pi^2} \left(\frac{F}{M}\right)^2 
h\left({F}/{M^2}\right),~~& 
(\Delta m^2_{{\tilde t}_R})_{\rm 1-loop}=  2 (\Delta m^2_{{\tilde t}_L})_{\rm 1-loop},\label{eq:1loop}
\end{align}
where the loop function reads $h(x)= x^2 + \frac{4}{5} x^4 + \mathcal{O}(x^6)$.
Note that a sizeable $A_t$ is an attractive feature of this model since it 
will permit to raise the Higgs mass to 126 GeV with moderate stop masses.

Moreover, there are additional two loop contributions to $\Delta_u,\Delta_d$ and to the other soft masses \cite{Evans:2011bea}:
\begin{align}
 & (\Delta^2_u)_{\rm 2-loop} = -\frac{9|\lambda_t|^2}{256 \pi^4}   |y_t|^2   \left(\frac{F}{M}\right)^2,~~ 
   (\Delta^2_d)_{\rm 2-loop} = -\frac{3|\lambda_t|^2}{256 \pi^4}  |y_b|^2   \left(\frac{F}{M}\right)^2,\nonumber\\
& (\Delta m^2_{{\tilde t}_L})_{\rm 2-loop}= \frac{|\lambda_t|^2}{128 \pi^4}\left(3|\lambda_t|^2+3|y_t|^2-\frac{8}{3}g_3^2-\frac{3}{2}g_2^2-\frac{13}{30}g_1^2\right) \left(\frac{F}{M}\right)^2,\nonumber\\
& (\Delta m^2_{{\tilde t}_R})_{\rm 2-loop}= \frac{|\lambda_t|^2}{128 \pi^4}\left(6|\lambda_t|^2+6|y_t|^2+|y_b|^2-\frac{16}{3}g_3^2-3 g_2^2-\frac{13}{15}g_1^2\right) \left(\frac{F}{M}\right)^2,\nonumber\\
& (\Delta m^2_{{\tilde b}_R})_{\rm 2-loop}= \frac{|\lambda_t|^2}{128 \pi^4} |y_b|^2 \left(\frac{F}{M}\right)^2.
\label{eq:2loop}
 \end{align}
\begin{figure}[t]
\begin{center}
  \includegraphics[width=0.465\textwidth]{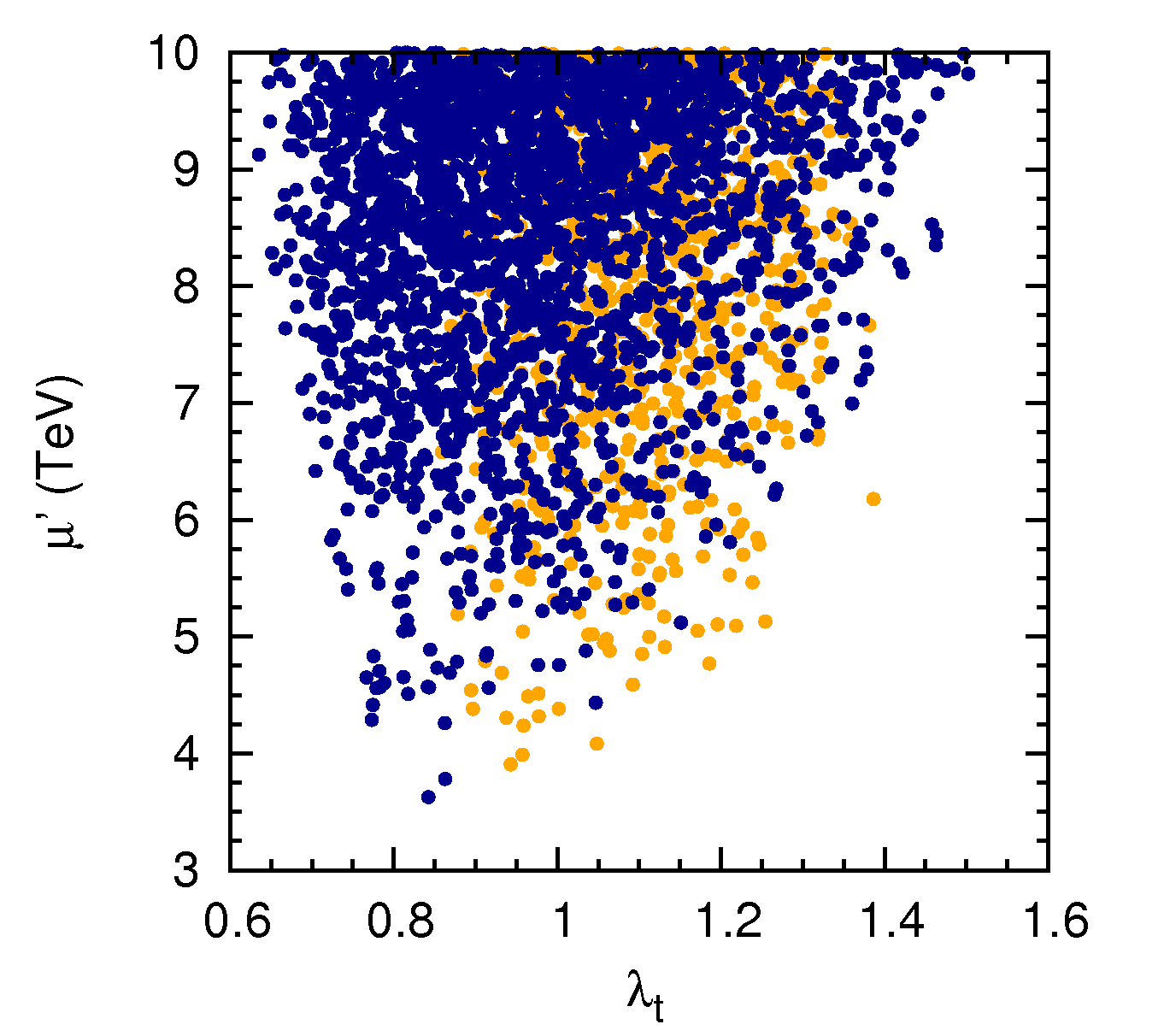}
  \includegraphics[width=0.48\textwidth]{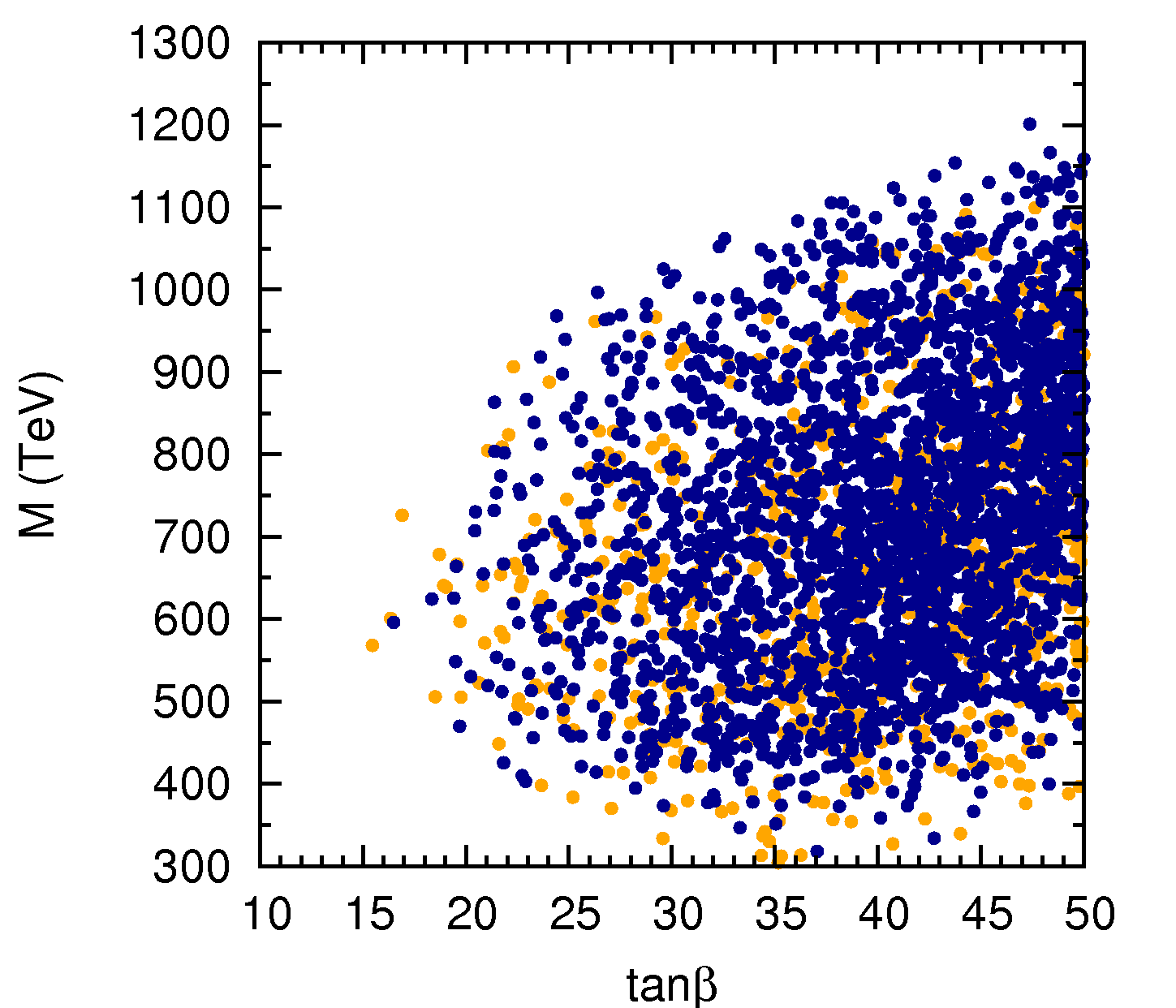}
\caption{\label{fig:yanaplots} Numerical scan for the model defined by Eq.~(\ref{summW}): all the displayed points feature a selectron NLSP and  $m_h = 126\pm 3$ GeV. See the text for details.} 
\end{center}
\end{figure}

We now 
perform a numerical study of the model in 
Eq.~\eqref{summW}
in order to illustrate that we indeed can obtain a selectron NLSP.
Setting the boundary conditions shown in Eqs.~(\ref{eq:tree}-\ref{eq:2loop}) on top of the ordinary GM contributions,
we employ {\tt SOFTSUSY 3.3.9} \cite{Allanach:2001kg} to run the soft terms from the messenger scale $M$ to the weak scale and to compute the low-energy spectrum.

We scan over the following ranges of the parameters:
\begin{align}
100~{\rm TeV} \le \Lambda\equiv\frac{F}{M} \le 500~{\rm TeV},& \qquad 2\times \Lambda \le M \le 5000~{\rm TeV}, \nonumber\\ 
5 \le \tan\beta  \le 50,& \qquad N=3 , \nonumber\\ 
0 <\mu'< 10 ~{\rm TeV},&\qquad 0 \le \lambda_t \le 2. 
\end{align}
Notice that we focus on low mediation scales, so that the deflection of Eq.~\eqref{eq:tree} is sizable, and we choose
three copies of messengers in order to increase the gaugino masses relative to the sfermion masses and avoid neutralino NLSP. Nevertheless, we assume, for simplicity, that only one messenger is coupled to the matter superfields.
We keep only the points that feature  a selectron NLSP at low energy, with a $\widetilde{\tau}_1$-$\widetilde{\ell}_R$ mass splitting larger than 20 GeV.
Moreover, we filter out solutions where any of the superpartners is heavier than 10 TeV.  We also impose the current bounds on the CP-odd Higgs mass \cite{CMS-PAS-HIG-12-050},
as well as 123 GeV $\le m_h \le$ 129 GeV. 
The result is shown in Figure \ref{fig:yanaplots}. The blue points also respect 
the absolute vacuum stability bound on the stop sector \cite{Claudson:1983et}:
\begin{equation}
 A_t^2 + 3\mu^2 \le 3(m^2_{\tilde{t}_L}+m^2_{\tilde{t}_R}).
\end{equation}
As we can see from Figure \ref{fig:yanaplots}, this additional (conservative) constraint has only a small impact on our parameter space.

\begin{figure}[t]
\begin{center}
  \includegraphics[width=0.5\textwidth]{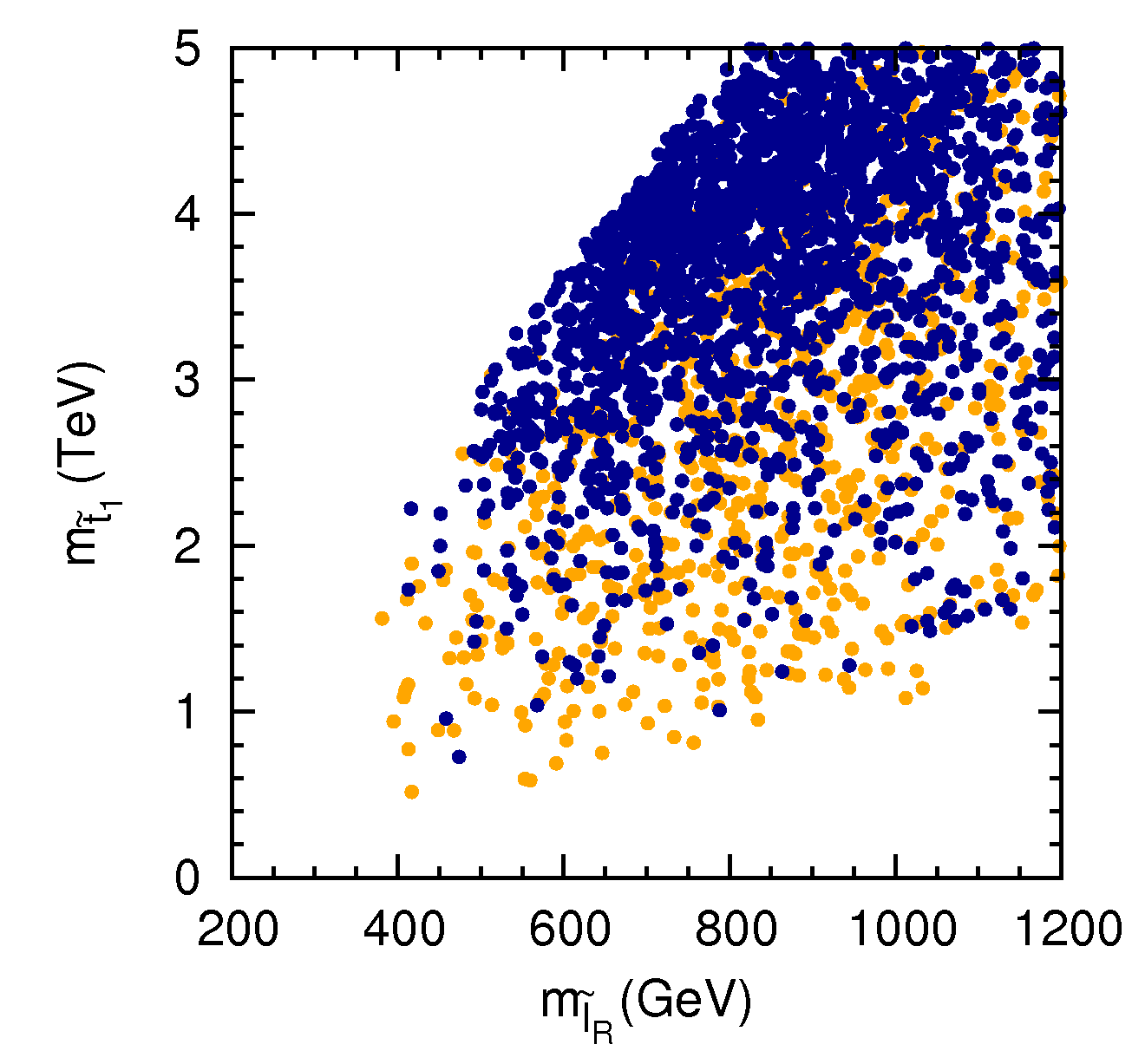}
\caption{\label{fig:yanaplot2} RH slepton vs.~stop masses for the model defined in Eq.~(\ref{summW}): all the displayed points feature a selectron NLSP and $m_h = 126\pm 3$ GeV. See the text for details.} 
\end{center}
\end{figure}

Comparing the two plots of Figure \ref{fig:yanaplots}, we can conclude that a selectron NLSP can be a quite generic feature of the model of Eq.~(\ref{summW}), under the following conditions:
\begin{itemize}
 \item a sizeable mixing parameter $\mu'$, larger than roughly $3$ TeV;
 \item $\lambda_t = \mathcal{O}(1)$, in order to generate a large $A_t$ and avoid problems with the CP-odd Higgs mass, as discussed in section \ref{sec:NLSP};
 \item moderate to large values of $\tan\beta$; 
 \item a low mediation scale: $M\lesssim 1200$ TeV.
\end{itemize}
The left panel of Figure \ref{fig:yanaplots} shows a quite definite range of $\lambda_t$ giving 
a selectron NLSP. This region is sharply bounded on the left by a too small $A_t$, giving $m^2_A<0$,
and on the right by a too large negative deflection of $m^2_{H_u}$ (c.f.~\eqref{eq:2loop}), implying
values of $\mu$ large enough to wash out our effect.

In Figure \ref{fig:yanaplot2}, we show the slepton NLSP and stop masses for the same scan.
As we can see, we can have a spectrum compatible with a 126 GeV Higgs mass with stops as light 
as $\approx$500 GeV. This is a consequence of the large A-term and the negative one-loop contributions
to the stop masses that are particularly effective for low values of $M$, cf.~Eq.~\eqref{eq:1loop} (for a detailed discussion see \cite{Calibbi:2013mka}).
The selectron/smuon co-NLSP can be as light as $380$ GeV, with the stau heavier but not decoupled
from the collider phenomenology.
Finally, let us remark that the small values needed for $M$ imply that this is a 
concrete and complete realization of promptly decaying 
selectron/smuon co-NLSP in GM.
We will discuss the related phenomenology in the following section. 

\section{LHC phenomenology}\label{sec:collider}
\begin{figure}[t]
\begin{center}
\includegraphics[scale=0.55]{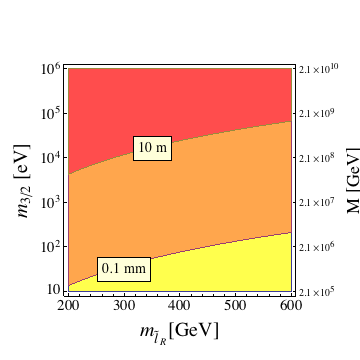}
\hspace{20pt}
\includegraphics[scale=0.45]{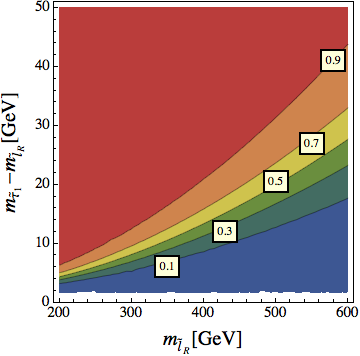}
\caption{On the left: the average distance the NLSP slepton travel before it decays. The yellow region corresponds to a prompt decay, $c\tau<0.1\,$mm, while the red region corresponds to the long-lived case, $c\tau>10\,$m. We display also the corresponding values of the messenger scale $M$ for $\Lambda=2\times10^5\text{ GeV}$. On the right:  $BR(\widetilde{\tau}_1\to\tau\ell\widetilde{\ell}_R)$ as a function of the RH slepton mass and the mass splitting between the stau and the sleptons.
}
\label{fig:plots} 
\end{center}
\end{figure}

In this section we discuss the bounds and phenomenology of GM scenarios in which the RH selectron and smuon are mass-degenerate co-NLSP. As can be seen in the left panel of \mbox{Figure \ref{fig:plots}}, depending on the mass of the gravitino and the sleptons, the slepton decay can be prompt or long-lived on collider time scales, or, in the intermediate case, it can give rise to a charged track that ends with displaced lepton vertex. 
Since the corresponding experimental searches and strategies are  different, we consider these three cases separately in the following subsections.

The NLSP 2-body decay width is given by the following universal formula (see for instance \cite{Martin:1997ns})
\be
\Gamma ( \widetilde{\ell}_R \to \ell \tilde G )=\frac{m_{\widetilde{\ell}_R}^5}{48 \pi m_{3/2}^2 M_{Pl}^2} \left(1-\frac{m_{\ell}^2}{m_{\widetilde{\ell}_R}^2}  \right)^4~.
\ee
The gravitino mass is most sensitive to the largest SUSY breaking scale and it is approximately given by 
\begin{equation}
m_{3/2} \simeq \frac{M}{\sqrt{3} M_{Pl}} \text{max}( \Lambda_{S_i}, \Lambda_{G_i} )\ ,
\end{equation}
where we are taking the SUSY breaking scale of the full hidden sector to be equal to the SUSY breaking scale of the messenger sector, as in direct mediation models.
For typical values of the SUSY breaking parameters, i.e.~the $\Lambda$'s, the decay length depends on the messenger scale $M$, as discussed in the previous sections. 
On the right vertical axes of Figure \ref{fig:plots} (left), we show the corresponding messenger scale for a given gravitino mass, having fixed  $\text{max} (\Lambda_{S_i}, \Lambda_{G_i}) =2 \times 10^5$ GeV. From this plot it is clear that all the GGM models give rise to either displaced or the long-lived decays, since they need a sufficiently long RG running. A promptly decaying NLSP is expected only for values of $M$ smaller than about $10^6$ GeV, which can easily be realized in the model of Section \ref{sec:Yanagida}.

\subsection{Prompt decays}

Since the RH sleptons are co-NLSP, their only decay channel is the 2-body decay to the gravitino, i.e.~$BR(\tilde\ell_R\to\ell \tilde G)=1$. These sleptons can be pair-produced at the LHC via Drell-Yan, $pp\to Z/\gamma\to\tilde\ell_R\tilde\ell_R$, and from searches in the final states $\ell^+\ell^- +\MET$ their mass is bounded to be $m_{\tilde\ell_{R}}>245$ GeV~\cite{Aad:2014vma}. Thus, this sets a lower bound on the entire spectrum in any GM model that realises the selectron NLSP scenario.  

The next superpartner in the spectrum above the RH sleptons is typically the lightest stau mass eigenstate $\tilde\tau_1$. The stau has two possible decay channels, either the 2-body decay to the gravitino, $\tilde\tau_1\to\tau\tilde G$, or the 3-body decay via an off-shell Bino, $\widetilde{\tau}_1\to\tau\ell\widetilde{\ell}_R$. Which one of these two decays that dominates depends on the masses of the stau, the sleptons, the gravitino and the Bino. In the right panel of \mbox{Figure \ref{fig:plots}} we show the contours of $BR(\widetilde{\tau}_1\to\tau\ell\widetilde{\ell}_R)$ as a function of the slepton mass and the mass difference between the stau and the sleptons. In this figure, the Bino mass is taken to be twice the slepton mass and always above the stau mass. The gravitino mass is set to be 20 eV, which is a value in the range where the decay of the co-NLSP sleptons to the gravitino is prompt.    

As can be seen in the right panel of Figure \ref{fig:plots}, it is only in the region where the mass splitting between the stau and the sleptons is very small that the 2-body decay of the stau to the gravitino dominates. In this region, the experimental bound on the stau mass is still the one set by LEP, $m_{\widetilde{\tau}_1}>87$~GeV \cite{Abbiendi:2005gc}, as the LHC searches for two hadronically decaying taus$+\MET$ are not yet sensitive \cite{ATLAS:2013yla}. Of course, in the scenario under consideration in which the RH sleptons are co-NLSP, this stau mass bound is irrelevant since the sleptons are bounded to be above 245 GeV.   

Let us now discuss the case where the dominant decay mode of the stau is the 3-body decay, i.e.~$\widetilde{\tau}_1\to\tau\ell\widetilde{\ell}_R\to \tau\ell\ell\tilde{G}$. Pair production of staus at the LHC, $pp\to Z/\gamma\to\tilde\tau_1\tilde\tau_1$, then gives rise to the final states $2\tau+4\ell+\MET$. Hence, we can use the LHC searches for 4$\ell+\MET$ to set bounds on the stau mass in this case. As can be seen from the right panel of \mbox{Figure \ref{fig:plots}}, in the case where the slepton mass is in the region around 245-400 GeV, it is enough with a stau-slepton mass splitting of around 10-20 GeV in order for the 3-body decay to dominate. For such small mass splittings the phase space will be suppressed and the two taus and the two leptons arising from the stau 3-body decay will be soft. In this compressed case, the efficiency for the lepton reconstruction drops and the 4$\ell+\MET$ searches loose sensitivity. Note that, since the sleptons must be at least 245 GeV, there will generically be two hard leptons and a significant amount of $\MET$ in the events. Further note that, since the taus can decay leptonically, the signal events can involve as many as 6 leptons.

In the ATLAS analysis \cite{ATLAS:2013qla}, based on 20.7 fb${}^{-1}$ of data at $\sqrt{s}=8$ TeV, a search is performed in final states with at least 4 leptons, each with $p_T>10$ GeV and $|\eta|<2.47/2.4$ for electrons/muons, and with $\MET>50$ GeV. There is no veto on taus or jets but there is a $Z$ veto rejecting all events in which the invariant mass of any pair, triplets or quadruplets of leptons is inside an interval of $\pm10$ GeV around the $Z$ boson mass. They set an exclusion bound at the 95\% CL on $\sigma\times\mathcal{A}\times\mathcal{\epsilon}<0.19$ fb, where $\mathcal{A}$ is the kinematic and geometric acceptance and $\epsilon$ is the detector efficiency. The trigger efficiency for the selected 4 lepton events is in the range of $90-100\%$, independently of the $p_T$ of the leptons.  Even though it is beyond the scope of this paper to perform a detailed analysis we see that in the range $\mathcal{A}\times\mathcal{\epsilon}=0.1-0.5$ 
one obtain a bound on the cross-section $\sigma(pp\to\tilde\tau_1\tilde\tau_1)<1.9-0.38$ fb, which translates into a mass bound on the stau of about $m_{\widetilde{\tau}_1}>245-340$ GeV. For lower values of $\mathcal{A}\times\mathcal{\epsilon}$, this search does not place a meaningful bound on the stau mass in the selectron NLSP scenario since the co-NLSP sleptons are already constrained from direct slepton pair production to be above 245 GeV.    

There is also a CMS analysis \cite{CMS:2013qda}, based on 19.5fb${}^{-1}$ of data at $\sqrt{s}=8$ TeV, in which a search is performed in final states with 4 leptons, where the leading lepton is required to have $p_T>20$ GeV, while the sub-leading leptons must have $p_T>10$ GeV. 
They set a 95\% CL exclusion bound on \mbox{$\sigma\times\mathcal{A}\times\mathcal{\epsilon}<0.17$ fb}, and therefore we expect that the bound on the stau mass in the selectron NLSP scenario will be comparable to the ATLAS bound discussed above set by the search \cite{ATLAS:2013qla}. For a discussion and comparison of the selectron NLSP scenario to the CMS search for events with three or more leptons \cite{Chatrchyan:2014aea}, see \cite{D'Hondt:2013ula}.

Which superpartner is above the lightest stau mass eigenstate is more model-dependent. In models like the one presented in Section \ref{sec:Yanagida}, where large top $A$-terms are generated, the lightest stop mass eigenstate $\tilde{t}_1$ can be the next superpartner in the spectrum. In the case where the stop mass is close to the slepton mass, due to the phase space suppression, the dominant decay mode will be the 2-body decay to a top+gravitino. The bound in this case is $m_{\tilde{t}_1}>740$ GeV \cite{CMS:2013cfa}. In the more generic case, where the mass splitting between the stop and the sleptons is greater than the mass of the top, the dominant decay mode will be the 3-body decay $\tilde{t}_1\to t \ell \widetilde{\ell}_R$, via an off-shell Bino. When this decay dominates, the pair produced stops give rise to the extremely clean final state $t\bar{t}+4\ell+\MET$. Even if we do not make use to the top-pair in this final state, and we simply apply the bound $\sigma\times\mathcal{A}\times\mathcal{\epsilon}<0.19$ fb from the ATLAS search in the final state with $4\ell+\MET$ \cite{ATLAS:2013qla}, for $\mathcal{A}\times\mathcal{\epsilon}=0.1$, the stop mass would be bounded to be above around 800 GeV. Clearly, the bound on the stop mass would be significantly stronger if also the presence of a top pair would be required.    

Other possible SM superpartners that can be light enough to be relevant at the LHC are the left-handed sleptons, $\widetilde{\ell}_L, \widetilde{\tau}_2, \widetilde{\nu}_{\ell,\tau}$. If they are accessible, they can be pair produced, $pp\to \widetilde{\ell}_L\widetilde{\ell}_L,\widetilde{\tau}_2\widetilde{\tau}_2,\widetilde{\nu}_{\ell,\tau}\widetilde{\nu}_{\ell,\tau},\widetilde{\ell}_L\widetilde{\nu}_{\ell},\widetilde{\tau}_2\widetilde{\nu}_{\tau}$. In the case where their dominant decay channel is the 3-body decay via an off-shell Bino, these processes give rise to final states involving 6, 5 or 4 leptons+$\MET$. As is clear from this discussion, for any superpartner that is accessible and whose dominantly decay channel is via an off-shell (or on-shell) Bino, the final state will involve at least 4 leptons+$\MET$.

\subsection{Long-lived sleptons}

In the case where the RH selectron/smuon co-NLSP decays outside of the detector, it can be reconstructed as a charged track, due to the energy released by ionisation, as it passes through the detector. Such a long-lived charged particle, appearing as heavy muon, has been searched for both by the ATLAS and the CMS collaborations. The ATLAS analysis \cite{ATLAS-CONF-2013-058} is based on 15.9 fb${}^{-1}$ of data at $\sqrt{s}=8$ TeV and the CMS analysis \cite{Chatrchyan:2013oca} is based on 5.0 fb${}^{-1}$ of data at $\sqrt{s}=7$ TeV and 18.8 fb${}^{-1}$ of data at $\sqrt{s}=8$ TeV. 

Both analyses consider pair production of long-lived NLSP staus. In the case where all SM superpartners, except for the lightest stau mass eigenstate, are decoupled, ATLAS/CMS sets a bound at $m_{\widetilde{\tau}_1}>267/339$ GeV. The (stronger) bound set by CMS, corresponds to a bound on the cross-section $\sigma(pp\to \widetilde{\tau}_1\widetilde{\tau}_1)$ at around 0.33 fb. If we translate this into the selectron NLSP scenario, in the case where the only contribution to this final state comes from the pair production of the mass-degenerate slepton co-NLSP, we simply divide the cross-section bound by a factor of 2 and get that the co-NLSP sleptons are bounded to be above around 400 GeV. If the stau is close to the sleptons, the stau provides an additional contribution to the final state, either by being long-lived itself or by decaying to the long-lived co-NLSP sleptons. In the case where the stau is nearly mass-degenerate with the sleptons, the three slepton families are bounded to be above around 435 GeV. 

In summary, in the case where the selecton/smuon co-NLSP decays outside the detector, $400$ GeV is a lower bound on the entire spectrum in any realization of the selectron NLSP scenario.

\subsection{Charged tracks with displaced vertices}

In the intermediate case where the NLSP selectron/smuon decay is non-prompt but takes place inside the detector volume, it may appear as a charged particle that decays, with a vertex that is separated from the original collision point, into an electron/muon (and $\MET$ carried by the invisible gravitino). If the emitted electron/muon is not reconstructed, the signature would be a charged track that at some point disappears inside the detector. As can be seen from Figure \ref{fig:plots} (left), the case where the NLSP gives rise to a displaced vertex is rather generic in this class of models.  

ATLAS has performed a search for such ``disappearing tracks" \cite{Aad:2013yna}, where the best efficiency is obtained for particles with a decay length of more than 30 cm and less than around 1 m. Since this search involves a lepton veto as well as a jet requirement of at least one jet with $p_T>90$ GeV, it is not sensitive to our scenario, for which sensitivity would be gained by instead {\it requiring} (at least) one additional lepton, instead of a jet.  

In the CMS analysis \cite{Chatrchyan:2012jna}, they search for a heavy resonance which decays into two neutral particles which travel a macroscopic distance in the detector before they decay to two leptons. The  final state they search for consists of a pair of opposite sign leptons which originate from a vertex that is displaced from the nominal interaction point by less than 2 cm away from the beam and they set limits for decay lengths in the range $0.1-200$ cm. Of course, our signal process would be qualitatively different both since the intermediate particle that mediates the displaced lepton vertex is a charged particle and since there would be only one lepton originating from the displaced vertex.  

Following this discussion, we propose a search, targeted for the case where the selectron/smuon NLSP gives rise to a displaced vertex, which consists of a combination of a search for a disappearing charged track and a search for an associated displaced vertex from which an electron/muon originates. We leave the discussion concerning the optimization and viability of such a search for future work.

\section{Summary and conclusions}

We have discussed scenarios of gauge-mediated SUSY breaking featuring mass-degenerate
selectron and smuon co-NLSP. By studying the MSSM RGEs, together with the requirements of having a successful EWSB and a viable low-energy spectrum (including $m_h \approx 126$ GeV), allowed us to identify the conditions under which the selectron NLSP scenario can be achieved. The key ingredient was found to be tachyonic soft masses in the UV. 
We discussed the conditions for the local stability of the vacuum, i.e.~for avoiding tachyonic scalars at the EW scale.
In order to perform a more detailed analysis concerning possible further constraints imposed by the global (meta)stability of the EW vacuum, one would
need to specify a more complete setup, and this is beyond the scope of the general discussion presented here. 
Nevertheless, in Section \ref{sec:Yanagida}, where we presented the simplest concrete model of selectron NLSP,
we imposed the vacuum stability bound coming from the stop sector with large A-terms (for a general vacuum stability analysis see for example 
\cite{Casas:1995pd,Evans:2008zx}).
Concerning the UV theory, one may wonder if the presence of tachyonic soft masses could lead to
color and charge breaking (CCB) minima and if they could spoil the cosmological evolution of our Universe.
This has been discussed in \cite{Ellis:2008mc}. First of all, the existence of CCB vacua should be ascertain by specifying the complete UV theory. Secondly, their existence could in any case be acceptable if the cosmological scenario prefers to populate the
EWSB vacuum, and this depends on the particular cosmological model considered. In the selectron NLSP scenario, such requirements could possibly translate in some 
further constraint on the UV completions, in terms of messenger models, that we have presented in Section \ref{messmodel}.
We leave a detailed study of this interesting issue for future investigation.

We have shown that the selectron NLSP scenario can be obtained within the framework of General Gauge Mediation, with tachyonic slepton
doublets at the messenger scale, as well as in the generalizations of this framework, characterized by additional (negative) contributions to the Higgs soft terms. For each class of models, we discussed the model-building requirements
for having selectron NLSP and, as a proof of existence, we provided concrete examples of models of weakly coupled messenger sectors. As a highlight of this discussion, we have shown that a selectron NLSP can be a natural consequence
of models with Yukawa-like matter-messenger couplings, sometime referred to as ``Yukawa-deflected gauge mediation'' in
the literature. These kind of models have recently aroused interest, as they can feature sizeable A-terms and thus
be able to accommodate a 126 GeV Higgs mass with a relatively light SUSY spectrum - in particular 
$m_{{\tilde t}_1}< 1$ TeV - thus addressing the serious problem posed to the GMSB framework by the Higgs mass measurement.

Finally, we have discussed the collider phenomenology and the LHC bounds on the selectron NLSP scenario, showing
how searches for EW production of SUSY states performed by ATLAS and CMS already provide non-negligible 
constraints on this framework. The absolute lower bound on the NLSP mass, and thus
on the entire spectrum, is $m_{\widetilde{\ell}_R}\gtrsim 245~(400)$ GeV in the case of promptly-decaying
(collider-stable) mass-degenerate RH selectron/smuon. 
For the case of 
selectron/smuon NLSP with intermediate lifetime, we proposed a new LHC search 
for charged tracks ending with a (displaced) vertex from which
an electron/muon originates.
Being characterized by extremely clean signatures, such as multilepton final states or charged tracks,
the prospects are very promising for extensively probing the selectron NLSP scenario in the upcoming 
$\sqrt{s}$ = 13/14 TeV run at the LHC.

\section*{Acknowledgements}
We are grateful to Simon Knapen and David Shih for many discussions on gauge mediation phenomenology and for sharing a private Mathematica code.
We would like to thank Matt Dolan for providing us a modified version of SOFTSUSY to perform some consistency check of our numerical analysis.
We would also like to thank Riccardo Argurio and Steve Abel for interesting discussions. 
A.M. acknowledges funding by the Durham International Junior Research Fellowship. The research of C.P. is supported by the Swedish Research Council (VR) under the contract 637-2013-475. The research of C.P. and D.R. is supported in part by IISN-Belgium (conventions 4.4511.06, 4.4505.86 and 4.4514.08), by the ``Communaut\'e Fran\c{c}aise de Belgique" through the ARC program and by a ``Mandat d'Impulsion Scientifique" of the F.R.S.-FNRS. R.A. is a Senior Research Associate of the Fonds de la Recherche Scientifique--F.N.R.S. (Belgium).

\let\oldbibitem=\bibitem\renewcommand{\bibitem}{\filbreak\oldbibitem}
\bibliographystyle{JHEP}
\bibliography{biblio}

\end{document}